\documentclass[aps,prx,twocolumn,superscriptaddress,groupedaddress]{revtex4}
\usepackage[dvips,letterpaper,margin=0.5in,bottom=0.75in]{geometry} 
\usepackage[english]{babel}
\usepackage{epstopdf}
\usepackage{pdfpages}
\usepackage{float}
\usepackage{enumerate}

\usepackage{yfonts}

\usepackage{amsmath} 
\usepackage{dsfont}
\usepackage{cancel}
\usepackage{amsthm} 
\usepackage{amssymb}	
\usepackage{graphicx} 
\usepackage{empheq}
\usepackage[absolute,overlay]{textpos}
\usepackage{xcolor}

\DeclareMathOperator{\atantwo}{atan2}

\newcommand{\imag}[0]{\mathrm{i}}

\begin{document}

\title{Topological phase transitions at finite temperature}

\begin{abstract}
The ground states of noninteracting fermions in one-dimension with chiral symmetry form a class of topological band insulators, described by a topological invariant that can be related to the Zak phase. 
Recently, a generalization of this quantity to  mixed states -- known as the ensemble geometric phase (EGP) -- has emerged as a robust way to describe topology at non-zero temperature.
By using this quantity, we explore the nature of topology allowed for dissipation beyond a Lindblad description, to allow for coupling to external baths at finite temperatures.
We introduce two main aspects to the theory of mixed state topology.
First, we discover topological phase transitions as a function of the temperature $T$, manifesting as changes in winding number of the EGP accumulated over a closed loop in parameter space. 
We characterize the nature of these transitions and reveal that the corresponding non-equilibrium steady state at the transition can exhibit a nontrivial structure -- contrary to previous studies where it was found to be in a fully mixed state. 
Second, we demonstrate that the EGP itself becomes quantized when key symmetries are present, allowing it to be viewed as a topological marker which can undergo equilibrium topological transitions at non-zero temperatures. 
\end{abstract}

\date{\today}

\author{Paolo Molignini}
\affiliation{T.C.M. Group, Cavendish Laboratory, J.J. Thomson Avenue, Cambridge CB3 0HE, United Kingdom}

\author{Nigel Cooper}
\affiliation{T.C.M. Group, Cavendish Laboratory, J.J. Thomson Avenue, Cambridge CB3 0HE, United Kingdom}
\affiliation{Department of Physics and Astronomy, University of Florence, Via G. Sansone 1, 50019 Sesto Fiorentino, Italy}

\maketitle

\section{Introduction}

Topology has emerged as a central paradigm in the theory of quantum phase transitions beyond the established Landau formalism~\cite{Wen:1990}.
Whereas Landau-type phases are described by continuous and local order parameters~\cite{Landau,Miransky-book}, topological phase transitions are heralded by integer-valued invariants that characterize the ground state wave function~\cite{Thouless:1982, Wen:1989, HasanReview:2010, Qi:2006, ChiuReview:2016, WenReview:2017}. 
Topological classifications exist for strongly correlated interacting systems which exhibit so-called topological order~\cite{Halperin:1982,Arovas:1984,Halperin:1984,Niu85,Kalmeyer:1987,Wen:1989-2, Witten:1989, Wen:1991,Moore:1991,Wen:1993, Wen:1999,Bonderson:2011} reflecting
long-range entanglement~\cite{Kitaev:2003, Kitaev:2006, Levin:2006, XieChen:2010},  as initially developed from the discovery of the fractional quantum Hall effect~\cite{Tsui:1982, Laughlin:1983}. 
However there are also topological phases with short-range entanglement, including the spin-1 Haldane phase~\cite{Haldane:1983, Affleck:1988, Gu:2009}, and the range of topological insulators and superconductors of gapped free-fermion systems~\cite{Kane-Mele:2005,Kane-Mele:2005-2,Bernevig:2006,Xu:2006,Fu:2007,Moore:2007,Qi:2008,Schnyder:2008,Kitaev:2009,Bernevig-book,ChiuReview:2016}. 

In the case of free-fermion topological insulators and superconductors, a very important role is played by the presence of symmetries, which lead to a rich classification of symmetry-protected topological (SPT) phases~\cite{Kane-Mele:2005,Kane-Mele:2005-2,Bernevig:2006,Xu:2006,Fu:2007,Moore:2007,Qi:2008,Schnyder:2008,Kitaev:2009,Bernevig-book,ChiuReview:2016}.
For such systems, topological invariants are constructed from the ground state wave functions of the underlying noninteracting fermions, with symmetries enforcing their quantization. 
These invariants can be derived from quantized geometric phases~\cite{Thouless:1982, HasanReview:2010, Qi:2006, QiReview:2011, ChiuReview:2016} which reflect knots or twists in the ground state wave function.

More recently, the exploration of SPT insulators and superconductors has been extended to more complex landscapes, such as systems with interactions~\cite{Rachel:2018,
Chen:2018, Chen-Sigrist-book:2019, MoligniniReview:2019, Zegarra:2019} or driven out of equilibrium~\cite{Lindner:2011, Kitagawa:2010, Liu:2013, Cayssol:2013, Thakurathi:2013, Graf:2013, Rudner:2013, Farrell:2016, Harper:2017, Roy:2017, Yao:2017, Molignini:2017, Molignini:2018, Esin:2018, Molignini:2019, Seetharam:2019, Molignini:2020-multifrequency, Rudner:2020, Harper:2020, Molignini:2021, mcginley:PRB2019,mcginley:PRR2019}.
The question of extending concepts of topology to mixed states has also attracted particular interest~\cite{Garate:2013, Bardyn:2013, Saha:2014, Saha:2015, Budich:2015, Grusdt:2017, Monserrat:2016, Bhattacharya:2017, Bardyn:2018, Coser:2019, Goldstein:2019, Lu:2020,Shapourian:2021,Lieu:2020,Altland:2021,Ashida-review:2020,Bergholtz-Review:2021}, not only for states at thermal equilibrium, but also for systems with engineered dissipation that lead to nonequilibrium states.

For mixed states, various generalizations of geometric phases and corresponding topological invariants have been proposed.
One way to tackle this question is to describe the effect of the environment with effective non-Hermitian Hamiltonians~\cite{Gong-nonhermitian:2018, Shibata:2019, Minganti:2019, Bergholtz-Review:2021,Rahul:2022, Tsubota:2022},
though, this construction is not very general. 
For instance, it is not suitable for the description of thermal equilibrium states.
Other approaches have confronted the problem more head on, and have defined topological invariants directly from mixed states.
One example is the Uhlmann phase~\cite{Uhlmann:1986, ViyuelaPRL112:2014, ViyuelaPRL113:2014, Huang:2014, Kempkes:2016, Quelle:2016, Carollo:2017, Viyuela:2018}, a formal generalization of the Berry phase~\cite{Barry:1983, Berry:1984, Wilczek:1984}.
However, while this quantity does exhibit quantization at nonzero temperature, its topological nature has been disputed on the basis that its construction relies on the definition of a global gauge, which is always topologically trivial~\cite{Budich:2015}.
Furthermore, while in one dimension the Uhlmann phase of gapped ground states recovers the closed-system topological invariant known as Zak phase, its construction in two dimensions fails to give a consistent definition of geometric phase because its winding takes different values depending on directionality~\cite{Budich:2015, Huang:2014, ViyuelaPRL113:2014, Wawer:2021-2}.

Another, more promising concept is that of the so-called Ensemble Geometric Phase (EGP)~\cite{Budich:2015, Linzner:2016, Bardyn:2018, Mink:2019, Unanyan:2020, Wawer:2021-1, Wawer:2021-2, Wawer:2021-3, Wawer:2022, Huang:2022}, which can be regarded as the mixed-state extension of Resta's polarization~\cite{Resta:1998} -- a reformulation of the Zak phase in terms of the expectation value of a many-particle momentum-translation operator.
The EGP appears to be a very suitable candidate to extend topology to mixed states.
It naturally extends the definition of closed-system topological invariants by replacing ground-state expectation values with statistical averages computed via the density matrix; it can be defined also when interactions are present~\cite{Unanyan:2020}; it correctly recollects the expected quantization in two dimensions, leading to well-defined Chern numbers~\cite{Wawer:2021-2}; it is directly measurable because it is based on a many-body observable~\cite{Bardyn:2018,Wawer:2022}.

So far, topological quantization was studied through the \emph{winding} of the EGP under the cyclic variation of external parameters.
In all cases, the corresponding topological phase transitions, signalled by a changed in quantization, were found to always occur at infinite temperature or for systems in which at least one mode becomes fully mixed.
Here, we show that this is an artefact of the fact that previous studies involved either thermal systems or dissipative systems described by Lindblad master equations.

We go beyond previous treatments by employing the Redfield master equation, which generalizes the Lindblad master equation in a way that allows us to define fermionic baths at both nonzero temperature and tunable chemical potential, i.e. establishing baths within the grand canonical ensemble.
We then explore the full nature of mixed-state topology for a combination of unitary dynamics and local Markovian dissipation.
Within this framework, we introduce two new aspects to the theory of mixed state topology.
First, a new kind of topological phase transition in the EGP winding can occur at \emph{finite} temperature and for a correlated nonequilibrium steady state.
This transition is different from previously studied cases because the quantization jumps between two nonzero, inequivalent values.
Second, we prove that the EGP itself can become $\mathbf{Z}_2$-quantized at equilibrium when key symmetries are present, and in certain parameter regimes.
We also show that, by tuning the values of the hoppings in the Hamiltonian, it is  possible to generate a topological phase transition between the two quantized values.
This result further strengthens the connection between the behavior of the EGP in open systems and the known theory of topological phase transitions in closed systems.

The rest of this paper is structured as follows.
In Sec.~\ref{sec:model}, we introduce the model, the methods, and the measures used to describe mixed-state topology.
In Sec.~\ref{sec:top-pumping-results}, we present our results for the topological quantization of the EGP winding in a nonequilibrium steady state and the corresponding temperature-driven topological phase transition.
In Sec.~\ref{sec:finite-T-quant}, we discuss our results for the system at equilibrium and demonstrate the EGP quantization, also by means of two analytical calculations.
Finally, Sec.~\ref{sec:conclusions} summarizes our discussion and provides an outlook for future studies.

\section{Model}
\label{sec:model}

We consider a Su-Schrieffer-Heeger (SSH) chain with $L$ unit cells~\cite{SSH:1979,Heeger:1988}, described by the Hamiltonian
%
\begin{equation}
\mathcal{H}_{\mathcal{S}} = -\sum_{n=1}^{L-1} \left[ t f_{n,A}^{\dagger} f_{n,B} + t' f_{n+1,A}^{\dagger} f_{n,B} + h.c. \right],
\label{eq:Ham-SSH}
\end{equation}
The system is composed of two sublattices $A$ and $B$. 
The operators $f^{\dagger}_{n,I}$ ($f_{n,I}$) denote creation (annihilation) operators for fermions on the $I=A,B$ sublattice of unit cell $n$, and satisfy canonical anticommutation relations $\{ f_{\nu}, f_{\mu}^{\dagger} \} = \delta_{\nu \mu}$. 
The particles can hop between the two sublattices with hopping strengths $t$ (intracell) and $t'$ (intercell).

The SSH chain at half filling is a prototypical model for a  symmetry-protected topological insulator in one dimension in closed systems.
Because of the staggered hopping configuration, the model possesses chiral symmetry~\cite{Asboth-book, ChiuReview:2016}.
When $t' < t$, the system is topologically trivial.
When $t' > t$, instead, the system is in a topological phase and topologically protected modes appear at the end of the chain. 
Various proposals have been made to exploit the properties of the edge modes in different contexts, from constructing thermoelectric devices~\cite{Boehling:2018}, to performing quantum computations and processing~\cite{Boross:2019,Dangelis:2020}, to building waveguides of magnetic excitation~\cite{Go:2020}, and more.
Furthermore, this model has been realized in numerous different platforms, including Rydberg atoms~\cite{Leseluc:2019,Kanungo:2022}, ultracold atoms~\cite{Atala:2013,Meier:2016,Zhang:2018:AiP,CooperReview:2019}, polaritonic micropillars~\cite{St-Jean:2017}, and optomechanical devices~\cite{Youssefi:2021}.
The variety of SSH model applications despite its simplicity make it an excellent system with which to investigate  topology at non-zero temperature.

The two different topological phases of the SSH model can be distinguished by a topological invariant constructed as a winding number from the momentum-space Hamiltonian~\cite{ChiuReview:2016}.
For inversion-symmetric systems, a related way of classifying the topological phases is offered by the Zak phase~\cite{Zak:1989}.
A major advantage in using the Zak phase to describe topology is its physical interpretability; it is the generalization of the Berry phase to Bloch wave functions in solids, it can be naturally extended to the concept of non-Abelian Wilson loops in multiband cases~\cite{Rhim:2017}, and it is experimentally measurable~\cite{Atala:2013}.
Furthermore, it can also be rewritten in terms of a polarization $P$ as ~\cite{Resta:1998}
\begin{equation}
\phi_Z = 2\pi P = \Im \log \left< \psi_0 \right| T \left| \psi_0 \right>,
\label{eq:Zak-phase}
\end{equation}
where $\left| \psi_0 \right>$ is the ground state and
\begin{align}
T &\equiv \exp \left( \frac{2\pi \imag}{L} X \right) \nonumber \label{eq:T-op}
\\
X &\equiv \sum_n n f_{n,A}^{\dagger} f_{n,A} + (n + 1/2) f_{n,B}^{\dagger} f_{n,B}.
\end{align}
$X$ is the (center of mass) position operator and $T$ is its generalization that assures a well-defined operator also for periodic boundary conditions~\cite{Resta:1998}.
This is a form that can be extended more easily to mixed state topology.
Due to the chiral symmetry, the Zak phase can only acquire two discrete values.
For the definitions given, one finds that $\phi_Z=-\frac{\pi}{2}$ for the nontopological phase, and $\phi_Z = + \frac{\pi}{2}$ for the topological one.

\begin{figure}[h]
\centering
\includegraphics[width=\columnwidth]{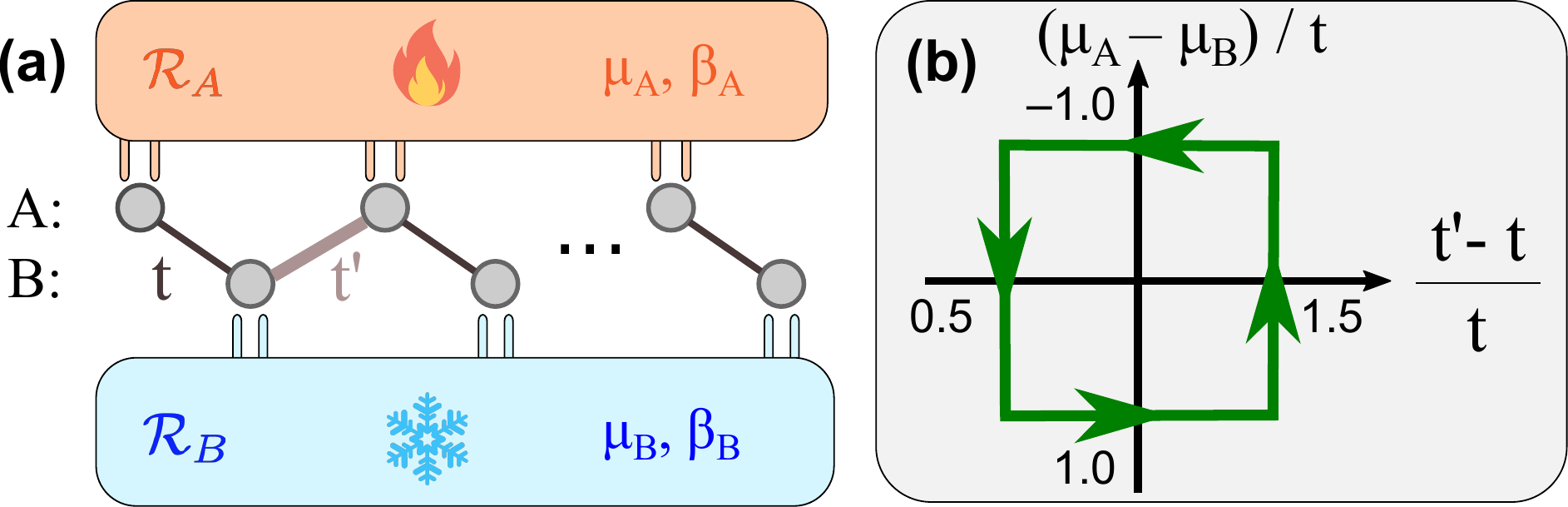}
\caption{(a) Sketch of the system analyzed in this article: a Su-Schrieffer-Heeger model with hoppings $t$ and $t'$ is attached to two Markovian fermionic reservoirs of chemical potential $\mu_i$ and inverse temperature $\beta_i$.
(b) Example of a pumping protocol performed in the parameter space spanned by the hoppings and the chemical potentials. The other pumping protocols described in this work are obtained by translating the path horizontally [see also inset in \ref{fig:pumping}(b)].
}
\label{fig:sketch}
\end{figure}

Since we are interested in describing the topology of this system at nonzero temperature and in non-equilibrium settings, we additionally couple it to two fermionic, Ohmic reservoirs $\mathcal{R}_{\text{A}}$ and $\mathcal{R}_{\text{B}}$.
Reservoir $\mathcal{R}_{\text{A}}$ ($\mathcal{R}_{\text{B}}$) is kept at chemical potentials $\mu_{\text{A}}$ ($\mu_{\text{B}}$) and inverse temperature $\beta_{\text{A}}$ ($\beta_{\text{B}}$), and both have a constant density of states.
In all our calculations we will set the Boltzmann constant to be $k_B=1$.
Furthermore, the two reservoirs couple differently to the system.
Reservoir $\mathcal{R}_{\text{A}}$ only couples to the $A$ sublattice, while reservoir $\mathcal{R}_{\text{B}}$ only couples to the $B$ sublattice.
A complete sketch of system and reservoirs is presented in Fig.~\ref{fig:sketch}(a)).

\subsection{Methods}

To describe the physics of the SSH model coupled to the two fermionic reservoirs, we employ the Redfield master equation (RME)~\cite{Redfield:1957, Breuer-Petruccione-book, Prosen:2010}.
The RME has several advantages over the more commonly used Lindblad master equation.
It is more general than the latter, because it retains oscillating terms that are otherwise ignored when the secular approximation is performed.
It contains only one generator for each system-bath interaction, and it allows to construct dissipation processes directly from the macroscopic state variables of the reservoirs, such as their temperatures and chemical potential.
This allows us to probe the effects of such state variables on the system, in particular with respect to its topology.
When it preserves positivity, the Redfield equation can also be more accurate than the Lindblad master equation~\cite{Mozgunov:2020}.

To describe the total system, we decompose its Hilbert space into a tensor product of the Hilbert subspaces of the system with Hamiltonian $\mathcal{H}_S$ and reservoirs with Hamiltonian $\mathcal{H}_{\mathcal{R}} \equiv \mathcal{H}_{\mathcal{R}_{\text{A}}} + \mathcal{H}_{\mathcal{R}_{\text{B}}} $.
The total Hamiltonian of the system can then be written in terms of such a tensor product as
\begin{equation}
\mathcal{H} = \mathcal{H}_{\mathcal{S}} \otimes \mathds{1}_{\mathcal{R}} + \mathds{1}_{\mathcal{S}} \otimes \mathcal{H}_{\mathcal{R}} + \lambda \sum_{\ell} X_{\ell} \otimes Y_{\ell},
\label{eq:full-Ham}
\end{equation}
where $\mathds{1}_{\mathcal{S}}$ ($\mathds{1}_{\mathcal{R}}$) indicates the identity operator on the Hilbert space of the system (reservoirs).
The last term in Eq. \eqref{eq:full-Ham} describes the interaction between system and reservoirs, with a coupling constant $\lambda$.
Throughout this work, we shall assume $\lambda$ to be small and equal for both reservoirs.
We remark that, while the value of $\lambda$ does impact transient dynamics and dictates the relaxation time to the steady state, it should not influence the behavior of the system at long times, which is the focus of our study.

The operators $X_{\ell}$ ($Y_{\ell}$) act on the Hilbert subspace of the system (reservoirs) and are chosen to be Hermitian and local. 
We shall consider baths that lead to the injection and removal of fermions for each site $(n,I)$ in the chain. 
We remark that this type of local dissipation differs from the nonlocal dissipation used in earlier works~\cite{Bardyn:2018}.
Motivated by exact solutions available for quadratic systems with Hermitian bath operators~\cite{Prosen:2010}, we cast them in terms of the Majorana operators
\begin{equation}
X_{\ell} \equiv X_{n, I, \alpha} = \begin{cases}
& \frac{1}{\sqrt{2}} \left( f_{n,I} + f_{n,I}^{\dagger}  \right), \quad \alpha=1 \\
& \frac{\imag}{\sqrt{2}} \left( f_{n,I} - f_{n,I}^{\dagger}  \right), \quad \alpha=2.
\end{cases}
\end{equation}

The RME is obtained by solving the Heisenberg equation of motion for the total system under the assumptions that the coupling between system and reservoir is weak ($\lambda$ small), that the initial density matrix is factorizable as $\rho_S(0) \otimes \rho_R(0)$, and that the bath correlation functions
\begin{equation}
\Gamma_{(j,I), (k,I)}^{\beta_I}(t) \equiv \lambda^2 \frac{\mathrm{Tr} \left[  
\tilde{Y}_j(t) Y_k e^{-\beta_{I} \mathcal{H}_{\mathcal{R}_{I}}}
\right]}{\mathrm{Tr} \left[ e^{-\beta_{I} \mathcal{H}_{\mathcal{R}_{I}}} \right]},
\label{eq:bath-corr}
\end{equation}
with $\tilde{Y}_j(t) \equiv e^{\imag t (\mathcal{H}_{\mathcal{R}_{I}} - \mu_I \mathcal{N}_{\mathcal{R}_{I}})} Y_j e^{-\imag t (\mathcal{H}_{\mathcal{R}_{I}} - \mu_{I} \mathcal{N}_{\mathcal{R}_{I}} )}$,
decay much faster than the time scale of the system dynamics (Born-Markov approximation)~\cite{Breuer-Petruccione-book, Prosen:2010}.
The RME so obtained then describes the Liouvillean dynamics $\hat{\mathcal{L}}$ of the system density matrix in terms of a coherent time evolution $\hat{\mathcal{C}}$ generated by the system Hamiltonian, and a dissipative part $\hat{\mathcal{D}}$ stemming from the interaction between system and reservoirs:
\begin{align}
\dot{\rho}(t) &\equiv \hat{\mathcal{L}}[\rho(t)] = \hat{\mathcal{C}}[\rho(t)] +  \hat{\mathcal{D}}[\rho(t)] \\
&= -\imag \left[ \mathcal{H}_{\mathcal{S}} , \rho(t) \right] \nonumber \\
& \quad + \sum_{j, k} \int_0^{\infty} \mathrm{d} \tau \: \Gamma_{k, j}^{\beta_{\mathcal{R}_{I}}}(\tau) \left[ e^{-\imag \tau \mathcal{H}_S} X_{j} e^{i \tau \mathcal{H}_S} \rho, X_{k} \right] + h.c.
\label{eq:rme}
\end{align}
Here, the dotted quantities indicate time derivatives, whereas the hats indicate superoperators acting in the Liouvillean space.

The bath correlation functions of Eq.~\eqref{eq:bath-corr} are more easily expressed in frequency space via a Fourier transform
\begin{equation}
\tilde{\Gamma}_{kj}^{\beta_{\mathcal{R}_{I}}}(\omega) = \int_{-\infty}^{\infty} \mathrm{d} \tau \: \Gamma_{k j}^{\beta_{\mathcal{R}_{I}}}(\tau) e^{-\imag \omega \tau}.
\end{equation}
The corresponding bath correlation spectral functions take the form
\begin{align}
\tilde{\Gamma}(\omega) &= \mathrm{diag}(\tilde{\Gamma}_A(\omega), \tilde{\Gamma}_B(\omega), \tilde{\Gamma}_A(\omega), \tilde{\Gamma}_B(\omega), \cdots) \\
\tilde{\Gamma}_I (\omega) &=  \lambda^2 g_I(\omega) \left[  f_+(\omega) \mathds{1} -  f_-(\omega) \sigma^y \right]
\end{align}
with $f_{\pm}(\omega) \equiv n_I(\omega) \pm (1- n_I(-\omega))$, $n_I(\omega) = \frac{1}{e^{\beta_I (\omega - \mu_I)} + 1}$ the Fermi-Dirac distribution of reservoir $\mathcal{R}_I$, and $g_I(\omega)$ its density of states.
As we will assume Ohmic baths of free fermions that couple equally to all unit cells throughout this work, we will set $g_I(\omega) \equiv \mathrm{const.}$.

\subsection{NESS observables}
\label{sec:NESS-obs}

We now address the question of how to solve the RME of Eq.~\eqref{eq:rme}.
We shall be focusing exclusively on the structure of the non-equilibrium steady state (NESS) in the RME, i.e. the behavior of the density matrix $\rho_{\mathrm{NESS}} \equiv \rho(t \to \infty)$.
In principle, this can be obtained by performing exact diagonalization of the full Liouvillean spectrum and then analyzing the eigenstate corresponding to the zero eigenvalue.
For large systems, this is typically a hard problem.
However, further simplifications can be performed when the Liouvillean is a quadratic form in the fermionic operators, i.e. when the Hamiltonian is quadratic and the bath operators are linear.
This is the case that we consider in the present work, which is best formulated in terms of Majorana operators
\begin{equation}
w_{2m-1} = f_{m} + f_m^{\dagger} \quad w_{2m} = \imag (f_m - f_m^{\dagger}),
\end{equation}
with the index $m$ now encompassing both the unit cell and sublattice indices, i.e. $m=(n,I)$.
In the Majorana representation, the Hamiltonian and the interaction operators can be written as
\begin{align}
\mathcal{H}_S &= \sum_{j,k=1}^{4L} w_j H_{jk} w_k \\
X_{\ell} &= \sum_{j=1}^{4L} x_{\ell,j} w_j,
\end{align}
with 
\begin{align}
H_{jk} &= \frac{\imag}{4} \left( \begin{array}{c c c c c c c c} 
0 & 0 & 0 & -t & 0 & 0 & 0 & \cdots \\
0 & 0 & t & 0 & 0 & 0 & 0 & \cdots \\
0 & -t & 0 & 0 & 0 & -t' & 0 & \cdots \\
t & 0 & 0 & 0 & t'  & 0 & 0 & \cdots \\
0 & 0 & 0 & -t' & 0  & 0 & 0  & \cdots \\
0 & 0 & t' & 0 & 0  & 0 & t & \cdots \\
0 & 0 & 0 & 0 & 0  & -t & 0 & \cdots \\
\vdots & \vdots & \vdots & \vdots & \vdots & \vdots & \vdots & \ddots 
\end{array} \right)
\end{align}
and $x_{\ell, j} = \delta_{\ell, j}$ for our particular choice of system and dissipation.

For such a quadratic problem, Prosen~\cite{Prosen:2008,Prosen:2010} showed that the Liouville space of $(2^{2L})^2$-dimensional operators, which the density matrix $\rho(t)$ is also a member of, has a Fock space structure which can be spanned by a set of $8L$ new Majorana operators.
The doubling of the space $4L \to 8L$ comes from assigning Majorana fermions to both bras and kets.
Each superoperator acting on the density matrix can then be rewritten as a quadratic form in such new operators, including the Liouvillean itself~\footnote{To be more precise, we consider only the projection of the Liouvillean onto the subspace composed of an even number of fermionic operators~\cite{Prosen:2010}.}.
By diagonalizing this quadratic form, it is possible to obtain an analytic expression for the two-point Majorana correlator in the NESS, $\left< w_j w_k \right>_{\text{NESS}} \equiv \mathrm{Tr} \left[ w_j w_k \rho_{\text{NESS}} \right]$.
This mathematical derivation is explained in more detail in appendix~\ref{app:Prosen-derivation}.

The analytic calculation of the two-point Majorana correlator in the NESS forms the base for the calculation of any other quantities.
By virtue of the quadratic form of the Liouvillean, a generalized Wick's theorem guarantees that all other NESS observables can be derived from it.
This includes the extension of the closed system topological invariants to nonzero temperature, which we will discuss next.

\subsection{Ensemble Geometric Phase}
\label{sec:EGP-Gauss}

After having defined the methods to calculate observables for the NESS of the open SSH chain, we now present the quantity that describes its topology.
We follow the approach defined in Refs.~\cite{Budich:2015, Linzner:2016, Bardyn:2018, Mink:2019, Unanyan:2020, Wawer:2021-1, Wawer:2021-2, Wawer:2021-3, Wawer:2022, Huang:2022},
where the Zak phase of Eq. \eqref{eq:Zak-phase} is naturally extended to mixed states by replacing the ground-state expectation value of the operator $T$ with its mixed-state analog:
\begin{equation}
\phi_E \equiv \Im \log \mathrm{Tr} \left[ \rho T \right].
\end{equation}
This generalized topological invariant is termed \emph{Ensemble Geometric Phase} (EGP).
To measure it, direct interferometric methods have been proposed~\cite{Bardyn:2018}, as well as indirect methods by means of coupling the original system to ancillary ones~\cite{Wawer:2022}.

Earlier studies on purely thermal or purely nonlocal dissipative systems~\cite{Bardyn:2018} have highlighted the topological character of the EGP.
Specifically, the winding of the EGP along a closed parameter cycle is quantized.
The quantized value depends on the path taken: it is nonzero when the cycle encircles gap-closing points of the Liouvillean, and zero otherwise.
This feature is analogous to what happens in topological pumping procedures in closed systems~\cite{Rice:1982, Thouless:1983, Asboth-book}.
Later studies~\cite{Unanyan:2020} have also highlighted that the quantization of the winding survives when interactions are present. 
However, in all previous studies the corresponding topological phase transition between different quantized values was always found to occur only at infinite temperature or when the state becomes fully mixed. 
In the following, we will show that it is actually possible to obtain topological transitions in the EGP also at \emph{finite} temperatures and for a correlated NESS.

To calculate the EGP analytically, we can again take advantage of the fact that the Liouvillean is a quadratic form in the Majorana operators.
The NESS can therefore be mapped to a Gaussian state described by Grassmann variables~\cite{Bravyi:2005,Bardyn:2013}.
We emphasize that this construction is not restricted to the particular dissipative SSH model studied here, but can be applied to any quadratic Liouvillean.
By recasting the problem in the language of Grassmann variables, we are able to rewrite $U \equiv \mathrm{Tr}\left[ \rho_{\mathrm{NESS}} T \right] / \mathrm{Tr}\left[ \rho_{\mathrm{NESS}} \right] $ as Gaussian integrals and use the rules of Grassmann calculus to obtain the following analytic expression in terms of pfaffians of matrices:
\begin{equation}
U = c \: \Omega \: \mathrm{Pf}(C)  \left[ \mathrm{Pf}(K_1 - C^{-1}) - \mathrm{Pf}(K_2 - C^{-1}) \right].
\label{eq:U-def}
\end{equation}
In this expression, $c=\exp \left(\frac{ \imag \pi}{2} (2N+3)\right)$ and $\Omega \equiv \prod_{k=1, k\neq N-1}^{2N} \cos \left( \frac{\pi(k+1)}{2N} \right)$ are constants. 
The covariance matrix $C_{jk} = \frac{\imag}{2} \mathrm{Tr} \left( \rho_{\text{NESS}}[w_j, w_k] \right)$ is the representation of $\rho_{\text{NESS}}$ as Gaussian state of Grassmann variables, and can be obtained from the two-point correlators via the procedure explained in section \ref{sec:NESS-obs}.
The matrices $K_1 = \bigotimes_{k=1, k \neq N-1}^{2N} \sigma_y^k \tan \left( \frac{\pi(k+1)}{2N} \right) \otimes \sigma_y^{N-1}$, and $K_2 = \bigotimes_{k=1, k \neq N-1}^{2N} \sigma_y^k \tan \left( \frac{\pi(k+1)}{2N} \right)$ define instead the Grassmann representation of the operator $T$ and are block diagonal.
The full derivation of formula \eqref{eq:U-def} and all its quantities is presented in appendix~\ref{app:grassmann-rep}, and forms the basis of our analysis of mixed state topology in the next sections. 
From it, the EGP can be easily obtained by calculating either directly the complex phase of $U$ (equilibrium case), or its winding number as $U$ traces a closed path on the complex plane when parameters are adiabatically modulated (nonequilibrium case).

\section{Finite-temperature topological pumping}
\label{sec:top-pumping-results}

We now present the results obtained by analyzing the SSH chain coupled to fermionic reservoirs within the RME formalism.
We begin by discussing how the EGP behaves in pumping procedures similar to those investigated in Refs.~\cite{Bardyn:2018} and~\cite{Unanyan:2020}.
We take inspiration from well-known topological pumping procedures that take place in 1D models in closed settings.

When the chiral symmetry is broken, for instance by adding a staggered on-site potential $u$ to the SSH model, the Zak phase loses its quantization.
The resulting model is often denoted as Rice-Mele model~\cite{Rice:1982}.
While the Zak phase is no longer quantized, it is possible to retrieve another quantization by adiabatically varying the parameters $(t'-t)/t$ and $u/t$ in time along a closed cycle, for instance described by the angle $\phi \equiv \atantwo (u/t')$.
Then the quantity $\Delta \phi_Z \equiv \frac{1}{2\pi} \oint \mathrm{d} \phi \partial_{\phi} \phi_Z$ is integer quantized and is associated with the number of particles (charges) pumped across the chain per cycle.
The quantized differential changes of the Zak phase are topological because they can be regarded as a two-dimensional Chern number when time is interpreted as a quasimomentum in an additional spatial dimension.
The quantization is nonzero only if the path encircles the gap closing point at $t'=t$, $u=0$. 
As a consequence, one can realize topological transitions between phases with different values of $\Delta \phi_Z$ ($\Delta \phi_Z = \pm 1 \leftrightarrow 0$) by shifting the path in parameter space until it crosses the gap closing point.

A similar reasoning has been studied before for open SSH chains that are purely dissipative, and where the role of tunneling is replaced by non-local dissipators that act on two neighbouring sites.
Here we study an open SSH chain that has both coherent tunneling and {\it local} dissipation, which we will employ to break inversion symmetry.
This can be achieved by having baths with different chemical potentials, i.e. $\mu_{\text{A}} \neq \mu_{\text{B}}$ in general.
In other words, the quantity $\Delta \mu \equiv \mu_{\text{A}} - \mu_{\text{B}}$ plays the open-system role of the staggered potential $u$ used in the closed system Rice-Mele model.
Similar to the closed system counterpart, we also define a closed loop in the parameter space spanned by $\Delta \mu / t$ and $(t'-t)/t$.
For simplicity, we employ the piecewise straight path illustrated in Fig.~\ref{fig:sketch}(b), but we expect our results to remain valid for other closed loops in parameter space.

Following the analogy with the closed-system topological pumping scenario, our expectation is that differential changes of the EGP accumulated along the path $\mathcal{P}$ will be topologically quantized,
\begin{equation}
\Delta \phi_E \equiv \frac{1}{2\pi} \oint_{\mathcal{P}} \left( \frac{u}{u^2 + v^2} \mathrm{d} v - \frac{v}{u^2 + v^2} \mathrm{d} u \right),
\end{equation}
where in this case we have just written the accumulated differential changes in the EGP as the winding number of the quantity $U= u + i v$ defined in \eqref{eq:U-def} along the path $\mathcal{P}$.
As we will see, not only is this realized, but the quantization can even change as a function of the inverse temperature $\beta$ and the shift $\delta$.

\begin{figure}[h!]
\includegraphics[width=\columnwidth]{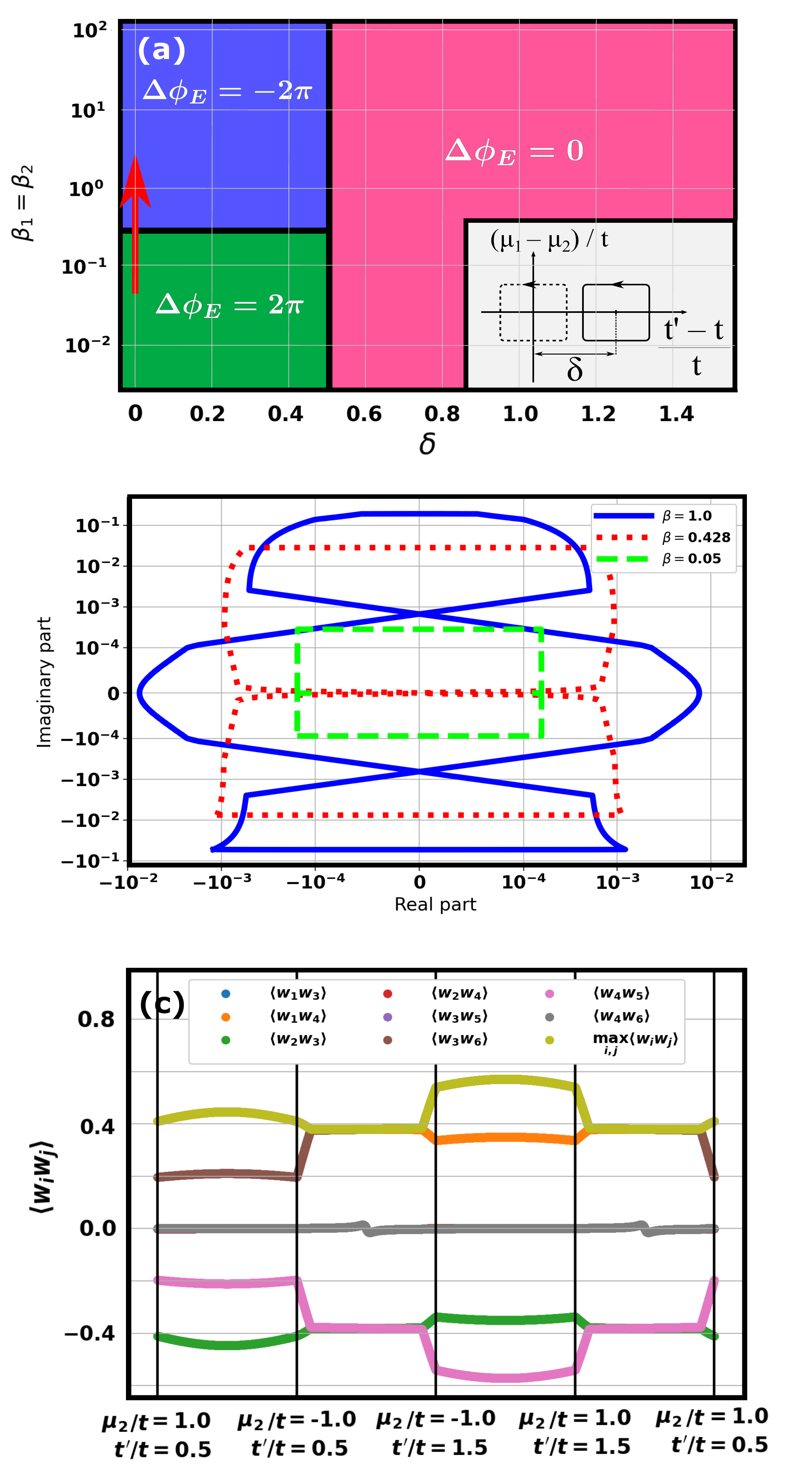}
\caption{
(a) Topological phase diagram for a $L=8$ system, defined by the integrated difference in the EGP, $\Delta \phi_E$, showing a temperature-driven topological phase transition.
The parameters of the phase diagram are the horizontal displacement of the path center $\delta$ (shown in the inset) and the inverse temperature $\beta = \beta_{\text{A}} = \beta_{\text{B}}$.
The path is constructed by varying $t'$ from $0.5$ to $1.5$ and $\mu_{\text{B}}$ from $-1.0$ to $1.0$, while all other parameters are kept fixed to $t=1.0$ and $\mu_{\text{A}}=0.0$.
(b) Illustration of the quantity $U$, whose winding is $\Delta \phi_E$, along the path traced in Fig. \ref{fig:sketch}(b) for various values of increasing temperature $\beta$ as indicated by the red arrow in the upper panel.
At a critical value around $\beta_c \approx 0.3$ (red dotted line), the winding number of $U$ jumps abruptly from $1$ to $-1$ as $U$ develops foldings which cross at the origin.
(c) Majorana correlators $\left< w_i w_j \right>$ along the path traced by $U$ for $\beta=1.0$ (the behavior for other values of $\beta$ is equivalent).
}
\label{fig:pumping}
\end{figure}

Fig.~\ref{fig:pumping} summarizes the results of our pumping procedure at nonzero temperature.
Panel (a) illustrates a topological phase diagram, where the value of $\Delta \phi_E$ is plotted as a function of a horizontal shift $\delta$ in the adiabatic cycle (depicted in the inset), and the inverse temperature of the reservoirs $\beta = \beta_{\text{A}} = \beta_{\text{B}}$.
We reiterate that while the reservoirs are kept at the same temperature, the system is in a nonequilibrium state because the chemical potentials are varied.
From the topological phase diagram, we can recognize three inequivalent regions where $\Delta \phi_E$ takes different discrete values.
The topological phases are separated by topological phase transitions occurring both as a function of $\delta$ and $\beta$.
We emphasize that throughout the adiabatic cycle, the NESS remains a correlated state, as we can see from Fig.~\ref{fig:pumping}(c).

At low values of $\beta$ and $\delta$, such that the path encircles the inversion-preserving, gap closing point at $t'=t$, $\mu_{\text{A}}=\mu_{\text{B}}$, we find $\Delta \phi_E= 2\pi$. 
This is the region (depicted in green in the figure) that is adiabatically connected to the quantized value of the EGP winding at infinite temperature found in earlier studies~\cite{Bardyn:2013,Unanyan:2020}.
We note that this quantization is preserved upon lowering the temperature by many orders of magnitude.
When the displacement is increased beyond $\delta_c = 0.5$, an abrupt jump to $\Delta \phi_E = 0$ occurs. 
This happens because the path crosses the gap-closing point and stops encircling it, and mirrors the situation known in the closed-system Rice-Mele model~\cite{Rice:1982}.
The phase at $\delta > \delta_c$ (depicted in pink in the figure) is analogous to the trivial phase discussed in Refs.~\cite{Bardyn:2013,Unanyan:2020}.

The most intriguing phase transition occurs however for $\delta < \delta_c$, as $\beta$ is increased.
At a critical value of $\beta = \beta_c \approx 0.3t$ the value of $\Delta \phi_E$ abruptly jumps from $2\pi$ to $-2\pi$.
This topological phase transition is of a new and different kind than the one triggered by the change in $\delta$.
This can be justified by considering both its intrinsic jump by \emph{two} integers, and the behavior of the quantity $U$ from which $\Delta \phi_E$ is calculated.
The latter is illustrated in panel (b) of Fig.~\ref{fig:pumping}.

In the region at low $\beta$, $U$ traces a closed, almost rectangularly-shaped loop around the origin in the complex plane (dashed green line).
Its winding depends on the direction in which we follow the path traced in parameter space.
For our choice of trajectory, $U$ winds in counterclockwise direction, and hence $\Delta \phi_E/(2\pi) = 1$.
A change in $\delta$ simply shifts the loop of $U$ away from the origin, eventually leading to $\Delta \phi_E = 0$.
When $\beta$ is increased, instead, the loop is gradually deformed around the origin.
Its vertical edges fold inward until they cross at the origin at $\beta = \beta_c$ (dotted red line), and move past one another for $\beta > \beta_c$ (solid blue line).
Because of the foldings, the path now winds in the opposite direction in the innermost loop, while the upper and lower loops do not contribute to the total winding number.
As a result, $\Delta \phi_E/(2\pi) = -1$.

It should be noted that previous studies had already shown the existence of the $\Delta \phi_E/(2\pi) = 1 \leftrightarrow 0$ topological phase transition in purely thermal systems~\cite{Bardyn:2013,Unanyan:2020}.
However, in these studies the transition always occurred at infinite temperature by going through a fully mixed state $\rho_{\text{NESS}} \propto \mathds{1}$.
Our case is remarkably different, because the topological phase transition occurs at finite temperature, i.e. $\beta \neq 0$ and through a NESS which remains correlated, as highlighted in Fig.~\ref{fig:pumping}(c).

\section{Finite-temperature topological quantization}
\label{sec:finite-T-quant}
 
We now focus on the equilibrium situation when both reservoirs are kept at the same inverse temperature and chemical potential, i.e. $\beta=\beta_{\text{A}}=\beta_{\text{B}}$ and $\mu = \mu_{\text{A}} = \mu_{\text{B}}$. 
In this case, the symmetry breaking between $A$ and $B$ sublattices does not occur.
We will show that under such conditions the EGP itself can become quantized in certain regimes, and that some form of quantization persists at all temperatures. 

\begin{figure}[h!]
\includegraphics[width=\columnwidth]{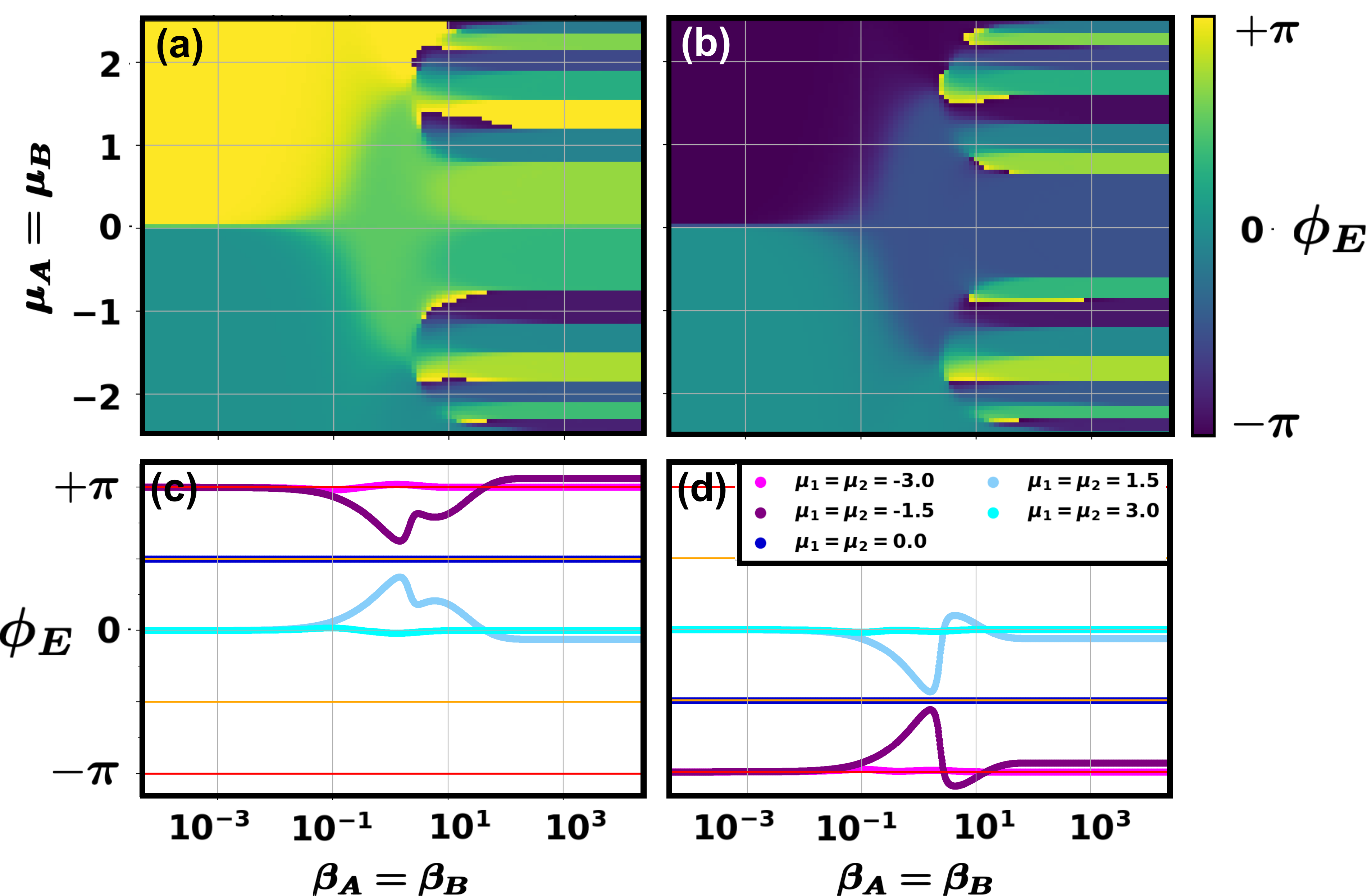}
\caption{EGP, $\phi_{E}$, at equilibrium as a function of the bath parameters $\beta_{\text{A}}=\beta_{\text{B}}$ and $\mu_{\text{A}}=\mu_{\text{B}}$ for a system of $L=8$ unit cells (OBC). 
 Upper panels: topological phase diagram for (a) $t' > t$ and (b) $t' < t$. 
Lower panels: cuts at fixed values of $\mu_A=\mu_B$, indicated in the legend, for (c) $t' > t$ and (d) $t' < t$.
}
\label{fig:pd-obc}
\end{figure}

We begin by describing the topological phase diagram that can be obtained by mapping $\phi_E$ as a function of $\beta$ and $\mu$, depicted in Fig.~\ref{fig:pd-obc} for both $t' > t$ (panel (a)) and $t' < t$ (panel (b)).
From this figure, we can distinguish three regimes:
\begin{enumerate}[(i)]
\item In the high-temperature limit, $\beta \to 0$, the EGP $\phi_E$ is quantized up to numerical accuracy at either $\phi_E = \pi$ (for $\mu > 0$) or $\phi_E = 0$ (for $\mu < 0$). 
An apparent topological transition between these two phases occurs at $\mu=0$.
Upon closer inspection, however, the value of the EGP $\phi_E$ at $\mu=0$ observed in the numerics is quantized at either $\pi/2$ (for $t'>t$) or $-\pi/2$  (for $t' < t$). 
This is illustrated in the bottom panels of Fig.~\ref{fig:pd-obc}.
\item At intermediate temperatures, the quantization at $\phi_E = \pi$ and $\phi_E = 0$ is lost. 
However, the $\phi_E = \pm \pi/2$ quantization at $\mu=0$ persists.
\item At low temperatures, the EGP $\phi_E$ assumes again a discrete set of values.
Both the number and the values of these discrete steps depend directly on the system size $L$.
As explained in appendix~\ref{app:filling}, this discretization is in one-to-one correspondence with the system filling.
In the limit $L \to \infty$, the discretization becomes a continuum of infinitely small steps, and can therefore not be topological in nature.
\end{enumerate}

Based on the observations extracted from the topological phase diagram, we now focus on the quantization observed in regimes (i) and (ii) and explain their physical origin.
First of all, because the EGP quantization at $\phi_E=0$ and $\phi_E=\pi$ is smoothly lost at intermediate values of the inverse temperature, it can only truly exist in the limit $\beta \to 0$.
In this limit, the NESS is a fully-mixed, infinite temperature state.
This is similar to the behavior observed in the earlier studies of topological pumping protocols mentioned in the earlier sections~\cite{Bardyn:2018,Unanyan:2020}.
In the $\beta \to 0$ regime, the EGP $\phi_E$ becomes a proxy for the average particle occupation in the chain, which is above half filling when $\mu >0$ and below it when $\mu < 0$.

\begin{figure}[h!]
\includegraphics[width=\columnwidth]{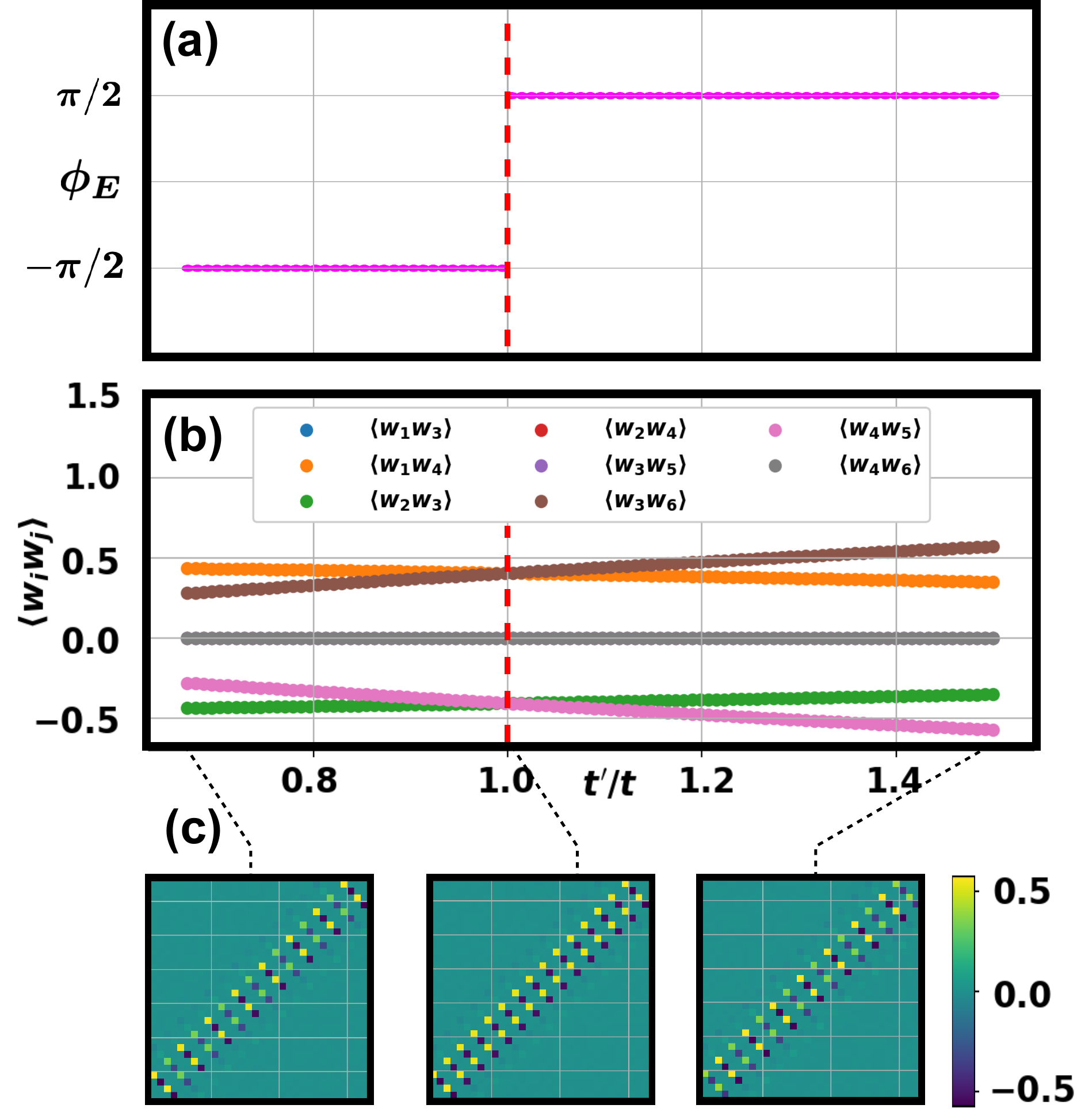}
\caption{(a) Finite-temperature transition in the EGP $\phi_E$ as a function of the hopping ratio $t'/t$. 
The EGP abruptly jumps from $-\frac{\pi}{2}$ to $\frac{\pi}{2}$.
The red dashed line indicates the transition at $t'=t$.
(b) The corresponding behavior of the Majorana correlators $\left<w_i w_j \right>$ between the first and second unit cell (the other unit cells behave in an analogous fashion).
(c) Illustration of the behavior of the full correlation matrix.
At the transition $t'=t$, the inter- and intra-cell correlations are equal, whereas for $t'>t$ ($t'<t$) the intercell (intracell) correlation dominates.}
\label{fig:EGP-maj-corr-equilibrium}
\end{figure}

\begin{figure}[h!]
\includegraphics[width=\columnwidth]{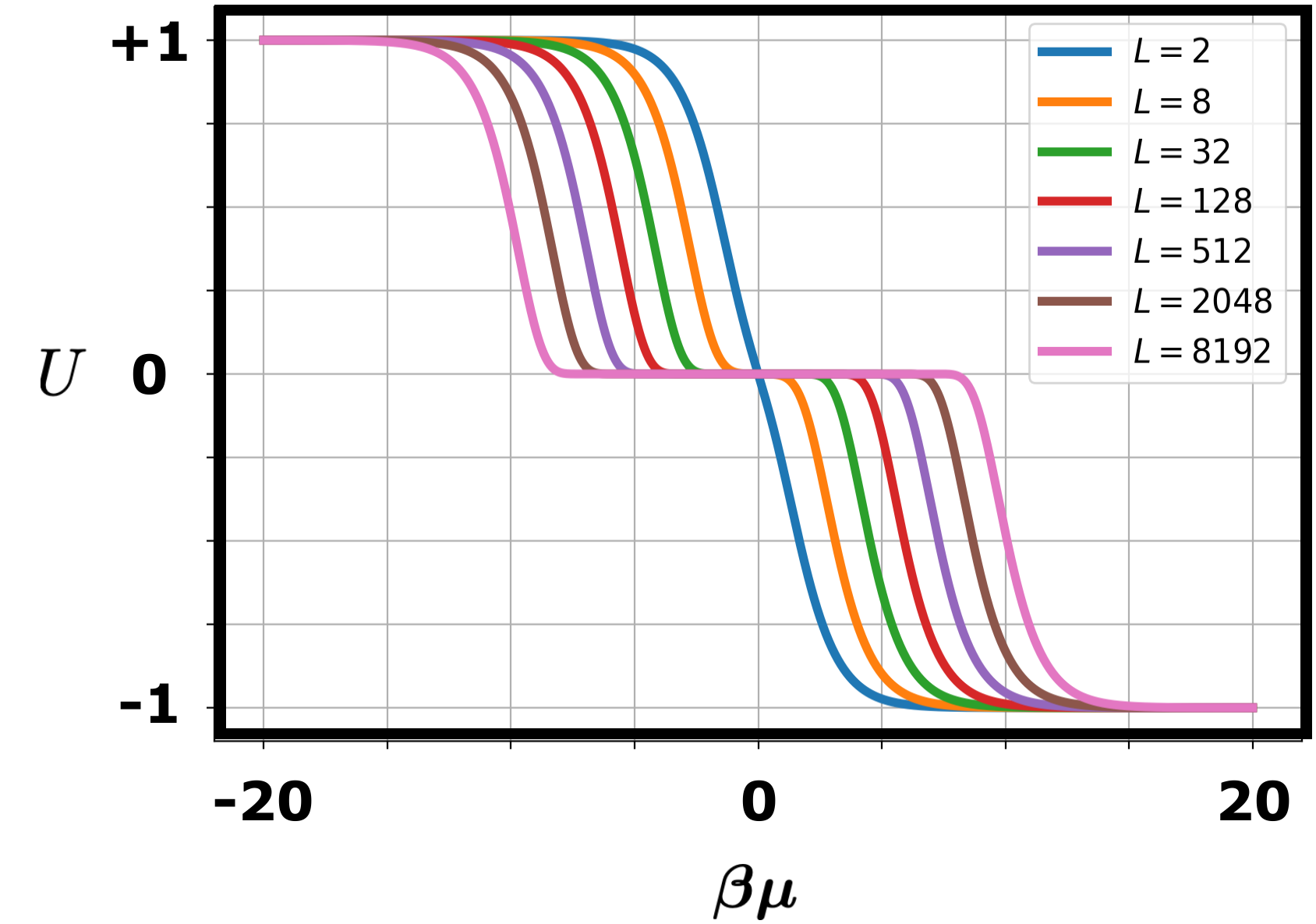}
\caption{Behavior of $U$ in the limit $\beta \to 0$, $\beta \mu \neq 0$ for various values of the system size $L$.
}
\label{fig:EGP-quantization-limit}
\end{figure}

The quantization of the EGP $\phi_{E}$ as $\beta \to 0$ can also be understood analytically.
In the long time limit, we expect the system to equilibrate with the reservoirs independently of initial conditions.
We can therefore describe it in the grand-canonical ensemble as a thermal Gibbs distribution $\rho$ described by inverse temperature $\beta$ and chemical potential $\mu$:
\begin{equation}
\rho \propto e^{-\beta (\mathcal{H} - \mu \mathcal{N})}
\end{equation}
where $\mathcal{H}$ is the Hamiltonian of Eq.~\eqref{eq:Ham-SSH} and $\mathcal{N} = \sum_{j=1}^L  \left( f_{j,A}^{\dagger} f_{j,A} + f_{j,B}^{\dagger} f_{j,B} \right) \equiv  \sum_{j=1}^L  \left( n_{j,A} + n_{j,B}\right)$ is the particle number operator.
In the $\beta \to 0$ limit, taking $\beta{\mathcal H}\to 0$ but allowing $\beta\mu {\mathcal N}$ to remain finite, the Gibbs distribution $\rho \to \exp(\beta\mu {\mathcal N})$. 
Writing $U\equiv U'/Z$ with $Z = {\mbox {Tr}}[\exp(\beta\mu {\mathcal N})]$ the partition function, we have
\begin{align}
U' &= \mathrm{Tr} \left[ {\rm e}^{\beta \mu {\mathcal N}} {\rm }e^{{\rm i} \frac{2\pi}{L} X} \right] \\
&= \sum_{\{n_{j,A}, n_{k,B} \}} \prod_{j,k} \left( e^{\beta \mu n_{j,A} + \imag \frac{2\pi}{L} j n_{j,A}} \right) \nonumber \\
& \qquad \qquad \qquad \times \left( e^{\beta \mu n_{k,B} + \imag \frac{2\pi}{L} (k+1/2) n_{k,B}} \right) \\
&= \prod_{j,k=1}^L \left( 1 + e^{\beta \mu + 2\pi {\rm  i} j /L} \right) \left( 1 + e^{\beta \mu + 2\pi {\rm i} (k+ 1/2) /L} \right)\,,
\label{eq:U-proof-quant}
\end{align}
where $\{ n_{j,A}, n_{k,B}\}$ stands for all possible configurations of fermionic occupations $n_{j,I}=0,1$.
%
This can be rewritten as the product
%
$U' = p(\eta) p(\xi)$
%
with $\eta \equiv - e^{\beta \mu}$ and $\xi \equiv -e^{\beta \mu} e^{{\rm i}\pi/L}$, by defining the function
\begin{align}
p(z) &\equiv \prod_{j=1}^L \left( 1 - q^j z \right),
\end{align}
with $q\equiv \exp(2\pi{\rm i}/L)$. The function $p(z)$ is a polynomial of degree $L$ in $z$, and its roots are evidently $z=1/q^j = e^{-2\pi {\rm i } j/L}$, i.e. the $L$-th roots of unity. Thus, by the fundamental theorem of algebra we can rewrite it as the equivalent polynomial
%
$p(z) = 1 - z^L$.
%
Using this fact, the expression for $U'$ simplifies to 
\begin{align}
U' &= (1 - \eta^L)(1 - \xi^L) \\
&= \left( 1 - (-1)^L e^{L \beta \mu} \right) \left( 1 + (-1)^L e^{L \beta \mu} \right) \\
&= 1 - e^{2 L \beta \mu}.
\end{align}
Adding the normalization $Z=\mathrm{Tr} \left[ \rho \right] = \mathrm{Tr} \left[ e^{\beta \mu {\mathcal N}} \right] = \prod_{j=1}^L (1 + e^{\beta \mu})^2$, we find
\begin{align}
U &= \frac{ 1 - {\rm e}^{2L \beta \mu}}{ \left(1 + {\rm e}^{\beta \mu} \right)^{2L}}.
\end{align}
This expression is real, which implies the quantization of the EGP to $\phi_E = 0,\pi$.
In the limit $\beta \mu \to \infty$, $U=-1$ and $\phi_E = \pi$, while in the limit $\beta \mu \to -\infty$, $U=+1$ and $\phi_E = 0$.
We therefore obtain an analytical expression that coincides with our numerical results in those limits.
The transition between phases $\phi_E = 0$ and $\phi_E = \pi$ occurs at $\beta\mu = 0$. 
However, one finds that over a range $\Delta (\beta \mu) \sim \ln(L)$ around $\beta\mu = 0$ the magnitude of $U$ is close to zero, rendering $\phi_E$ very difficult to measure close to the transition, cf. Fig.~\ref{fig:EGP-quantization-limit}.


Besides the transition in the infinite-temperature limit, the system exhibits another, more interesting transition for $\mu=0$ at any finite temperature, e.g. $\beta \in (0, \infty)$.
As the difference between the two hoppings $t$ and $t'$ is what determines the quantized value of the EGP at $\mu=0$, it should be then possible to realize a topological phase transition by tuning them.
This is indeed the case, as shown in panel (a) of Fig.~\ref{fig:EGP-maj-corr-equilibrium}: as $t'$ is varied and becomes larger than $t$, the EGP abruptly jumps from $\phi_E = -\pi/2$ to $\phi_E = + \pi/2$.
We stress that this is a topological phase transition that can occur at any finite and nonzero temperature.
The corresponding physical behavior of the system can be understood by examining the correlation between different Majorana sites, and is illustrated in panels \ref{fig:EGP-maj-corr-equilibrium}(b) and (c).
Throughout the whole transition, the NESS remains in a correlated state with nonzero values of $\left< w_j w_k \right>$.
Below the transition, $t'<t$, the intracell correlation dominates.
At the transition $t'=t$, the correlation becomes uniform across the whole chain.
Above the transition, $t'>t$, the situation is then reversed and the intercell correlation becomes stronger.
This behavior is very similar to what occurs in the closed-system SSH model as a function of the hopping strengths.
The EGP behaves exactly how the Zak phase would in the closed setting.
However, we stress that in our case the system is fully open and thermalized.

We now demonstrate analytically how this nonzero-temperature EGP quantization can be related to the system fulfilling certain symmetries. 
We restrict our analysis to the SSH chain with $L$ unit cells and periodic boundary conditions, but we believe that our findings should apply to any system that fulfils the same symmetry requirements.
We consider the following transformation:
\begin{equation}
f_{i,A} = - \tilde{f}_{i,A}^{\dagger}, \qquad f_{i,B} = \tilde{f}_{i,B}^{\dagger}.
\end{equation}
This transformation implicitly relies on the chiral symmetry of the model because it transforms $A$ and $B$ sublattice differently.
The operators appearing in the definition of $U$ are changed under this mapping as
\begin{align}
\tilde{\mathcal{H}} &= \mathcal{H} \\
\tilde{\mathcal{N}} &= 2L - \mathcal{N} \\
\tilde{X} &= L(L+1) + \frac{L}{2} - X \\
e^{\imag \frac{2\pi}{L} \tilde{X}} &= - e^{-\imag \frac{2\pi}{L} X}.
\end{align}
Then we can rewrite $U$ as
\begin{align}
 U &= \mathrm{Tr} \left[ - e^{-\beta \left( \tilde{\mathcal{H}} - 2 L \mu + \mu \tilde{N} \right)} e^{- \imag \frac{2\pi}{L} \tilde{X}} \right].
\end{align}
In particular, for $ \beta \mu= 0$, this simplifies to
\begin{align}
U &= -\mathrm{Tr} \left[ \rho e^{- \imag \frac{2\pi}{L} \tilde{X}} \right] = - U ^*,
\label{eq:U-transf}
\end{align}
since we can equally well take the trace over $f_{i, \alpha}$ or $\tilde{f}_{i, \alpha}$.
The EGP is the phase of $U$, and thus the restriction of Eq.~\eqref{eq:U-transf} imposed by the symmetries implies its quantization:
\begin{align}
e^{i \phi_{\mathrm{E}}} &= - e^{-i \phi_{\mathrm{E}}} \\
\Rightarrow \quad \phi_{\mathrm{E}} &= \frac{\pi}{2} \mod \pi.
\end{align}
We remark that, contrary to previous studies using the canonical ensemble in which the quantization of the EGP was predicted to occur only in the thermodynamic limit $L \to \infty$~\cite{Bardyn:2018}, our proof demonstrates the EGP quantization also for finite system size, provided the relevant symmetries are present.

\section{Conclusions and outlook}
\label{sec:conclusions}

We have shown how topological phase transitions can occur at finite temperatures in a one-dimensional open system.
Concretely, we have analyzed a Su-Schrieffer-Heeger model, a prototypical symmetry-protected topological insulator, coupled to two fermionic reservoirs and described in the formalism of the Redfield master equation.
Contrary to previous studies, our model requires only local dissipation combined with nearest-neighbor coherent tunneling.
The reservoirs were chosen such that each couples to only one of the two sublattices of the model.
To describe the topology of such an on open system, we have employed the Ensemble Geometric Phase (EGP), a many-body observable that naturally extends the notion of the Zak phase to mixed states.
We have calculated the Ensemble Geometric Phase of the steady state in two different scenarios.

First, we have analyzed the out-of-equilibrium behavior of the system when both reservoirs are kept at the same temperature, but their chemical potentials and the system hoppings are varied adiabatically in time along a closed loop.
We discovered that in this case the EGP is not quantized, but differential changes along the loop in parameter space are. 
This behavior is similar to what occurs for pumping procedures in closed systems and in other studies of mixed-state topology.
The quantization changes as a function of whether the loop encircles gap closing points or not.
More remarkably, changing the temperature also affects the quantization and leads to a temperature-driven topological phase transition.

Second, we have considered the system at thermal equilibrium, when both the temperature and the chemical potential are kept equal across the two reservoirs.
In this scenario the EGP itself is quantized and the quantized values depend on the hopping parameters of the system.
The quantization can occur either at infinite temperature for any chemical potential, or at finite temperature when the chemical potential is zero.
In this case, we proved the quantization analytically by leveraging the chiral symmetry of the problem.
By tuning the values of the hoppings, we showed that it is therefore possible to achieve a topological phase transition at finite temperature. 

Our study elucidates the untapped potential of extending concepts of topological phases and phase transitions to out-of-equilibrium and thermal systems.
Furthermore, it illustrates that temperature, long thought to be mainly a detrimental factor to topological quantization, can not only be compatible with it, but also induce topological phase transitions.
Nevertheless, the concept of symmetries appears to remain central also for finite-temperature topology.

Our work opens up a broad range of future directions of study.
These could include generalizing our results to generic quadratic systems without explicitly relying on a particular form of the Hamiltonian, but by considering the possible classes of the Altland-Zirnbauer symmetry classification.
The EGP quantization could also be explored in higher dimensions, still in conjunction with different symmetries.
Other studies of thermal systems have shown that differential changes of the EGP can be interpreted as the ``Chern number'' of a higher-dimensional system, similarly to what happens in closed systems
~\cite{Wawer:2021-3, Wawer:2022}.
However, for such systems no transition as a function of the temperature was found.
It would then be natural to ask whether a nonequilibrium construction with adiabatically changing chemical potentials as in our present work could instead lead to a transition in higher dimensions.
It would also be interesting to explore the connection between the EGP quantization and the recently discovered symmetry classifications of open topological systems~\cite{Lieu:2020, Altland:2021, Lieu:2022}.
Another direction of study could be exploring the connection between EGP quantization and the topology of effective non-Hermitian Hamiltonians derived from master equations~\cite{Minganti:2019,Ashida-review:2020,Bergholtz-Review:2021}.
On the more experimental front, a natural question to ask is whether the EGP quantization can be measured.
Ultracold atomic systems are the ideal arena for this endeavor, given the possibility of engineering tailored dissipation and the proposals of detecting the EGP in interferometric measurements~\cite{Bardyn:2018}.

\acknowledgments

We gratefully acknowledge funding from the ESPRC Grant no. EP/P009565/1 and a Simons Investigator Award.
We would like to thank Rosario Fazio, Tilman Esslinger, and Sebastian Diehl for useful discussions. 


\appendix

\section{Calculation of the two-point Majorana correlator in the NESS}
\label{app:Prosen-derivation}

We summarize here the analytic expression for the two-point Majorana correlator in the NESS, $\left< w_j w_k \right> \equiv \mathrm{Tr} \left[ w_j w_k \rho_{\text{NESS}} \right]$.
Our calculation follows Refs.~\cite{Prosen:2008, Prosen:2010}, where the Liouvillean superoperator of a quadratic system is shown to possess a decomposition
\begin{equation}
\hat{\mathcal{L}} = \hat{\mathbf{a}}^T A \hat{\mathbf{a}} - A_0 \hat{\mathds{1}}.
\end{equation}
in terms of new Majorana operators $\hat{a}_r$, $r=1, \dots, 8L$.
In this expression, $A$ is an $8L \times 8L$ complex antisymmetric matrix termed \emph{structure matrix}, and $A_0$ is a scalar.
They are expressed as
\begin{align}
A_{2j -1, 2k-1} &= - 2 \imag H_{jk} - M_{jk} + M_{kj} \\
A_{2j -1, 2k} &= \imag M_{kj} + \imag M^*_{jk} \\
A_{2j, 2k-1} &= - \imag M_{jk} - \imag M^*_{kj} \\
A_{2j, 2k} &= -2 \imag H_{jk} - M^*_{jk} + M_{kj} \\
A_0 &= \mathrm{Tr} M + \mathrm{Tr} M^*,
\end{align}
where $M$ is a \emph{bath matrix} that encodes the effect of the reservoirs and takes the form
\begin{align}
M &\equiv \sum_{j} \mathbf{x}_{j} \otimes \mathbf{z}_{j}
\end{align}
with
\begin{align}
\mathbf{z}_{j} &= \pi \sum_{k} \sum_{m=1}^{2L} \bigg[ 
\tilde{\Gamma}_{jk}^{\beta_{\mathbf{R}_i}}(-4\epsilon_m) (\mathbf{x}_k \cdot \mathbf{u}_m) \mathbf{u}^*_m  \nonumber \\
& \quad \qquad \qquad + 
\tilde{\Gamma}_{jk}^{\beta_{\mathbf{R}_i}}(4\epsilon_m)  (\mathbf{x}_k \cdot \mathbf{u}^*_m) \mathbf{u}_m
\bigg].
\end{align}
Here, $\epsilon_m$ and $\mathbf{u}_m$ are respectively the eigenvalues and eigenvectors of the system Hamiltonian, i.e. $H \mathbf{u}_m = \epsilon_m \mathbf{u}_m$ and $H \mathbf{u}^*_m = -\epsilon_m \mathbf{u}^*_m$ since the Hamiltonian in Majorana representation fulfils $H^* = - H$.
The structure matrix $A$ can be further diagonalized as $A=V^T \mathrm{diag} \{ \beta_{\text{A}}, -\beta_{\text{A}}, \cdots, \beta_{4L}, -\beta_{4L} \} J V$ with $VV^T=J$ and $J \equiv \sigma^x \otimes \mathds{1}_{4L}$.
We remark that in the numerics it is necessary to perform an additional Schmidt orthonormalization procedure to guarantee the condition $V V^T = J$ if degeneracies in the rapidity spectrum are present such as in the SSH model studies in this work~\cite{Dangel-thesis:2017, Dangel:2018}.
The two-point correlator is finally calculated in terms of the eigenvectors of the structure matrix as~\cite{Prosen:2010}
\begin{equation}
\left< w_j w_k \right>_{\mathrm{NESS}} = 
2 \sum_{m=1}^{4L} V_{2m, 2j-1} V_{2m-1,2k-1}.
\end{equation}
%


\section{EGP discretization at zero temperature}
\label{app:filling}

\begin{figure}[h]
\centering
\includegraphics[width=\columnwidth]{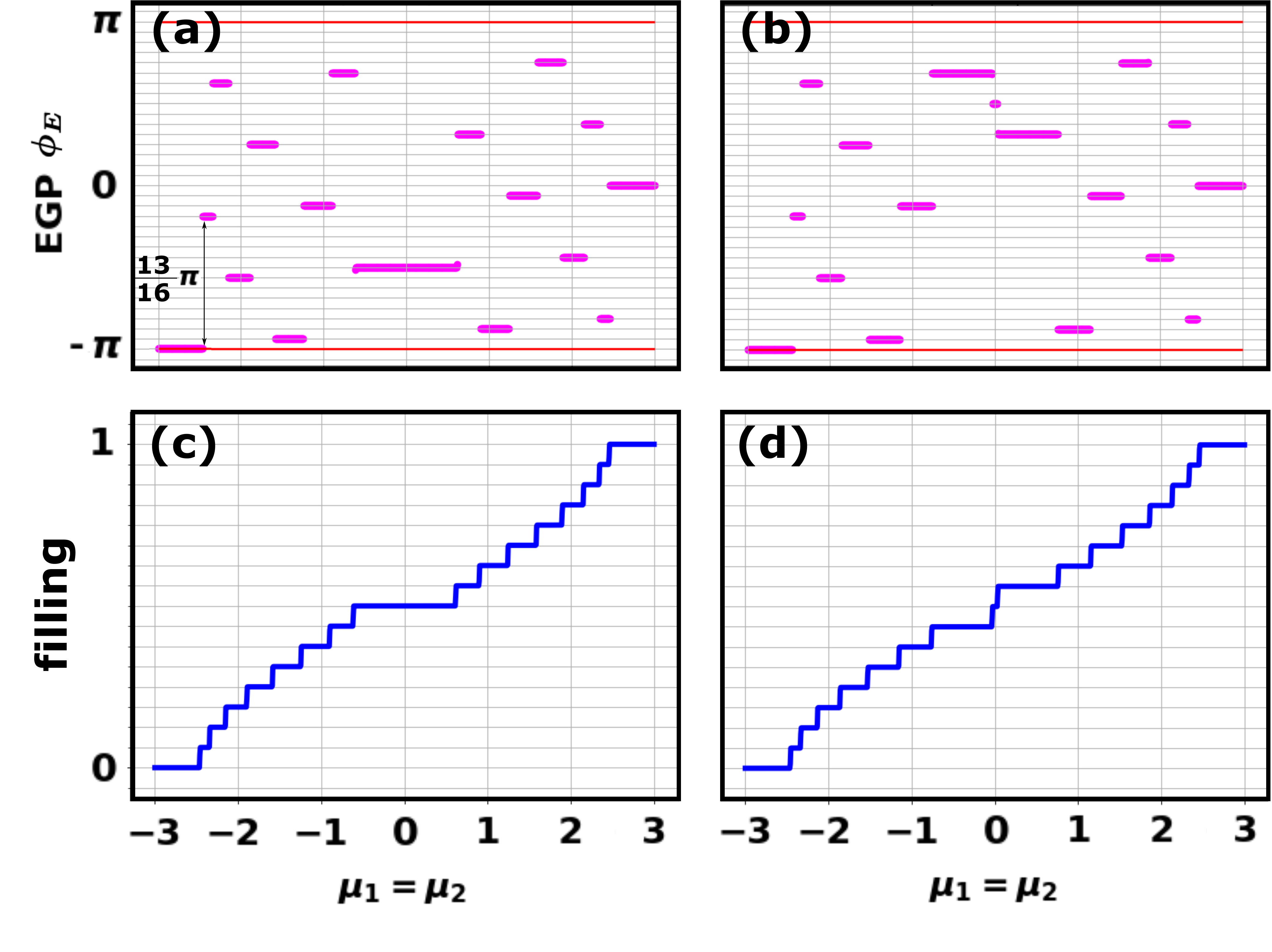}
\caption{
Behavior of a system with $L=8$ unit cells (OBC) in the $\beta \to \infty$ limit.
(a)-(b) EGP $\phi_E$ as a function of chemical potentials $\mu_1 = \mu_2$ for $t=1.5$, $t'=1.0$ (left panel) and $t=1.0$, $t'=1.5$ (right panel).
(c)-(d) Particle filling as a function of $\mu_1=\mu_2$ for the same parameters.
}
\label{fig:filling}
\end{figure}

In this appendix, we show results that explain the origin of the EGP quantization at equilibrium in the zero-temperature limit ($\beta \to \infty$).
In Fig.~\ref{fig:filling}, we plot the behavior of the EGP as a function of the chemical potential for two different regimes of $t'/t$, corresponding to a closed system (a) without and (b) with topological edge modes.
At first sight, the quantized values look random.
However, if we measure the jump in the EGP across two consecutive plateaus, we realize that this has a constant value of $\Delta \phi_E = \frac{2L-3}{2L} \pi \mod 2\pi$ ($2L$ is the total number of sites in the system).
This is the change in the EGP associated with a jump in the average filling factor of the fermionic chain.
We can see this by comparing the behavior of the EGP with the filling illustrated in Fig.~\ref{fig:filling}(c) and (d).
At zero temperature, as we increase the value of the chemical potential particles are pumped into the system one by one, and the average filling increases in steps of $1/2L$.
We remark that the presence or absence of the topological edge modes at half filling has an impact both on the behavior of the filling itself and on the EGP quantization.
In the limit of an infinitely long chain, where the modes are completely decoupled from each other, the filling jumps by two when crossing $\mu_1=\mu_2=0$.
At the same time, we would see a jump of $2\Delta \phi_E$ in the EGP.
For finite system sizes, as in the results shown here, a very small region around $\mu_1=\mu_2=0$ exists where the robust $\pi/2$ quantization still shows up in the numerics.

\section{Grassmann representation of the Ensemble Geometric Phase}
\label{app:grassmann-rep}

As explained in section \ref{sec:EGP-Gauss}, the EGP can be evaluated analytically by mapping the Majorana operators of the Liouvillean space to Grassmann variables, and then using known identities for Gaussian states.
In this appendix, we illustrate the steps that lead to this analytic results.

We begin by constructing a representation $\omega$ of products of Majorana operators in terms of Grassmann variables $\theta$ as~\cite{Bravyi:2005}
\begin{equation}
\omega(w_p w_q \cdots w_r, \theta) \equiv \theta_p \theta_q \cdots \theta_r, \qquad \omega(1,\theta) =1.
\end{equation}
This definition is then extended by linearity to arbitrary operators $X$ of the Clifford algebra: $X \mapsto \omega(X,\theta)$.
The Grassmann variables are anticommuting, such that
\begin{equation}
\{ \theta_i, \theta_j \} = 0 \qquad \theta_i^2 = 0.
\end{equation}
Because of this property, the operators appearing in the Liouvillean can be readily written as Gaussian forms (exponentials) of Grassmann variables, e.g.
\begin{equation}
f_1 f_1^{\dagger} = \frac{1}{2}(1 + \imag w_1 w_2) \mapsto \frac{1}{2}(1 + \imag \theta_1 \theta_2) = \frac{1}{2} \exp( \imag \theta_1 \theta_2 ).
\end{equation}
In particular, the NESS density matrix $\rho_{\text{NESS}}$ has the following Gaussian form~\cite{Bardyn:2013}
\begin{equation}
\omega(\rho_{\text{NESS}}, \theta) = \frac{1}{2^{2N}} \exp \left( \frac{\imag}{2} \theta^T C \theta \right),
\label{eq:Grassmann1} 
\end{equation}
where $\theta = \left(\theta_1, \cdots, \theta_{2N} \right)$ and $C_{jk} = \frac{\imag}{2} \mathrm{Tr} \left( \rho_{\mathrm{NESS}}[w_j, w_k] \right)$ is the covariance matrix that can be computed via third quantization as explained in section \ref{sec:NESS-obs}.

The matrix form of $\omega(T, \theta)$ can be obtained instead by evaluating the definition of the operator $T$ defined in Eq.~\eqref{eq:T-op}. 
In Majorana representation, we can write
\begin{equation}
T = c \prod_{k=1}^{2N} \left[ \sin \left( \frac{\pi(k+1)}{2N} \right) w_{2k-1} w_{2k} + \cos \left( \frac{\pi(k+1)}{2N} \right) \mathds{1} \right],
\end{equation}
with $c=\exp \left(\frac{\imag \pi}{2} (2N+3)\right)$.
The Grassmann representation $\omega(T, \theta)$ is obtained by simply replacing $w_i \mapsto \theta_i$.
Let us define $S(k) \equiv \sin \left( \frac{\pi(k+1)}{2N} \right)$, $C(k) \equiv \cos \left( \frac{\pi(k+1)}{2N} \right)$, and $T(k) \equiv \tan \left( \frac{\pi(k+1)}{2N} \right)$.
Then we can write, as $C(N-1)=0$ and $S(N-1)=1$,
\begin{align}
\omega(T, \theta) &= c \prod_{k=1}^{2N} \left( S(k) \theta_{2k-1} \theta_{2k} + C(k) \right) \\
& = c \prod_{k=1}^{N-2} \left(S(k) \theta_{2k-1} \theta_{2k} + C(k) \right) \nonumber \\
& \qquad  \times S(N-1) \theta_{2N-3} \theta_{2N-2} \nonumber \\
& \qquad  \times \prod_{k'=N}^{2N} \left( S(k') \theta_{2k'-1} \theta_{2k'} + C(k') \right) \\
& = c \prod_{k=1}^{N-2} C(k) \left( T(k) \theta_{2k-1} \theta_{2k} + \mathds{1} \right) \nonumber \\
& \qquad  \times \left( -1 + 1 \theta_{2N-3} \theta_{2N-2} \right)  \nonumber \\
& \qquad  \times \prod_{k'=N}^{2N} C(k') \left(T(k') \theta_{2k'-1} \theta_{2k'} + \mathds{1} \right)  \\
& = c \underbrace{\prod_{k=1, k\neq N-1}^{2N} C(k)}_{\equiv \Omega} \underbrace{\exp \left( \sum_{k=1}^{N-2} T(k) \theta_{2k-1} \theta_{2k} \right)}_{\equiv \alpha} \nonumber \\
& \qquad  \times \underbrace{\left( -1 + 1 + \theta_{2N-3} \theta_{2N-2} \right)}_{\equiv \exp(\theta_{2N-3} \theta_{2N-2}) - 1} \nonumber \\
& \qquad  \times \underbrace{\exp \left( \sum_{k'=N}^{2N}  T(k') \theta_{2k'-1} \theta_{2k'}\right)}_{\equiv \gamma} \\
&= c  \Omega ( \alpha \exp(\theta_{2N-3} \theta_{2N-2}) \gamma - \alpha \gamma) \\
&= c \Omega \left[ \exp \left( \frac{\imag}{2} \theta^T K_1 \theta \right) - \exp \left( \frac{\imag}{2} \theta^T K_2 \theta \right) \right],
\label{eq:Grassmann2}
\end{align}
where $\theta = \left( \theta_1, \cdots, \theta_{4N} \right)$, and $K_1$ and $K_2$ are the following matrices:
\begin{align}
K_1 &= \bigotimes_{k=1, k \neq N-1}^{2N} \sigma_y^k \: T(k) \otimes \sigma_y^{N-1} \\
K_2 &= \bigotimes_{k=1, k \neq N-1}^{2N} \sigma_y^k \: T(k).
\end{align}
Here, $\sigma_y^j$ is the Pauli matrix that spans a $2\times2$ space corresponding to site $j$.

Armed with the Grassmann representations of $\rho_{\text{NESS}}$ and $T$, we can now calculate
$\phi_E = \Im \log \mathrm{Tr} \left[ \rho_{\text{NESS}} T \right]$ explicitly.
We first note that the trace of two operators $X$, $Y$ living in the Clifford algebra of the Liouvillean have the following representation as Gaussian integral of Grassmann fields:
\begin{equation}
\mathrm{Tr} \left(X Y \right) = \left( -2 \right)^{2N} \int \mathrm{D} \theta \mathrm{D}\mu \: \exp \left( \theta^T \mu \right) \omega(X, \theta) \omega(Y, \mu),
\end{equation}
where $\int \mathrm{D} \theta \equiv \int \mathrm{d} \theta_N \cdots  \mathrm{d} \theta_2  \mathrm{d} \theta_1$, and similarly for $\mu$.
By employing the well-known identities for Gaussian integrals of Grassmann fields $\theta$, $\mu$,
\begin{align}
\int \mathrm{D} \theta \exp \left( \frac{\imag}{2} \theta^T C \theta \right) &= \imag^{2N} \mathrm{Pf} (C)
\label{eq:Gauss-int1} \\
\int \mathrm{D} \theta \exp \left( \eta^T \theta + \frac{\imag}{2} \theta^T C \theta \right) &= \imag^{2N} \mathrm{Pf} (C) \exp \left(-\frac{\imag}{2} \eta^T C^{-1} \eta \right),
\label{eq:Gauss-int2}
\end{align}
and the Grassmann representations \eqref{eq:Grassmann1} and \eqref{eq:Grassmann2} above, we can evaluate $\mathrm{Tr} \left[ \rho_{\mathrm{NESS}} T \right]$ explicitly.
To make the notation lighter, we focus only on the $K_1$ summand of Eq.~\eqref{eq:Grassmann2}.
The part for the summand containing $K_2$ can be obtained analogously.
\begin{align}
\mathrm{Tr} \left[ \rho_{\mathrm{NESS}} T \right] 
&= \frac{(-1)^{2N} c \Omega}{2} \int \mathrm{D} \theta \exp \left( \frac{\imag}{2} \theta^T K_1 \theta \right) \nonumber \\
& \qquad  \times \int \mathrm{D} \mu \exp \left(\theta^T \mu \right) \exp \left( \frac{\imag}{2} \mu^T C \mu \right) \\
&= \frac{(-1)^{2N} c \Omega}{2} \int \mathrm{D} \theta \exp \left( \frac{\imag}{2} \theta^T K_1 \theta \right) \nonumber \\
& \qquad \quad \times \int \mathrm{D} \mu \exp \left(\theta^T \mu + \frac{\imag}{2} \mu^T C \mu \right) \\
&= \frac{(-\imag)^{2N} c \Omega}{2} \mathrm{Pf}(C) \nonumber \\
& \qquad \quad \times \int \mathrm{D} \theta \exp \left( \frac{\imag}{2} \theta^T (K_1 - C^{-1}) \theta \right) \\
&= \frac{c \Omega}{2} \mathrm{Pf}(C) \mathrm{Pf}(K_1 - C^{-1}),
\end{align}
where in the second equivalence we have used $e^{A} e^{B} = e^{A+B}$ because $\theta$ and $\mu$ are independent fields and hence commute, 
in the third equivalence we have used Eq.~\eqref{eq:Gauss-int2}, and in the fourth equivalence we have used Eq.~\eqref{eq:Gauss-int1}.
The total expression is then
\begin{align}
\mathrm{Tr} \left[ \rho_{\mathrm{NESS}} T \right] &= c \Omega \mathrm{Pf}(C)  \Bigg( \mathrm{Pf}(K_1 - C^{-1}) \nonumber \\
& \qquad \qquad \qquad \qquad - \mathrm{Pf}(K_2 - C^{-1}) \Bigg).
\end{align}
%

\bibliography{many-body-biblio}

\begin{thebibliography}{138}
\expandafter\ifx\csname natexlab\endcsname\relax\def\natexlab#1{#1}\fi
\expandafter\ifx\csname bibnamefont\endcsname\relax
  \def\bibnamefont#1{#1}\fi
\expandafter\ifx\csname bibfnamefont\endcsname\relax
  \def\bibfnamefont#1{#1}\fi
\expandafter\ifx\csname citenamefont\endcsname\relax
  \def\citenamefont#1{#1}\fi
\expandafter\ifx\csname url\endcsname\relax
  \def\url#1{\texttt{#1}}\fi
\expandafter\ifx\csname urlprefix\endcsname\relax\def\urlprefix{URL }\fi
\providecommand{\bibinfo}[2]{#2}
\providecommand{\eprint}[2][]{\url{#2}}

\bibitem[{\citenamefont{Wen}(1990)}]{Wen:1990}
\bibinfo{author}{\bibfnamefont{X.-G.} \bibnamefont{Wen}},
  \bibinfo{journal}{Int. J. Mod. Phys. B} \textbf{\bibinfo{volume}{4}},
  \bibinfo{pages}{239} (\bibinfo{year}{1990}).

\bibitem[{\citenamefont{Landau}(1937)}]{Landau}
\bibinfo{author}{\bibfnamefont{L.~D.} \bibnamefont{Landau}},
  \bibinfo{journal}{Zh. Eksp. Teor. Fiz.} \textbf{\bibinfo{volume}{7}},
  \bibinfo{pages}{19} (\bibinfo{year}{1937}).

\bibitem[{\citenamefont{Miransky}(1994)}]{Miransky-book}
\bibinfo{author}{\bibfnamefont{V.~A.} \bibnamefont{Miransky}},
  \emph{\bibinfo{title}{Dynamical Symmetry Breaking in Quantum Field Theories}}
  (\bibinfo{publisher}{World Scientific Publishing Co.}, \bibinfo{year}{1994}).

\bibitem[{\citenamefont{Thouless et~al.}(1982)\citenamefont{Thouless, Kohmoto,
  Nightingale, and den Nijs}}]{Thouless:1982}
\bibinfo{author}{\bibfnamefont{D.~J.} \bibnamefont{Thouless}},
  \bibinfo{author}{\bibfnamefont{M.}~\bibnamefont{Kohmoto}},
  \bibinfo{author}{\bibfnamefont{M.~P.} \bibnamefont{Nightingale}},
  \bibnamefont{and} \bibinfo{author}{\bibfnamefont{M.}~\bibnamefont{den Nijs}},
  \bibinfo{journal}{Phys. Rev. Lett.} \textbf{\bibinfo{volume}{49}},
  \bibinfo{pages}{405} (\bibinfo{year}{1982}).

\bibitem[{\citenamefont{Wen}(1989)}]{Wen:1989}
\bibinfo{author}{\bibfnamefont{X.-G.} \bibnamefont{Wen}},
  \bibinfo{journal}{Phys. Rev. B} \textbf{\bibinfo{volume}{40}},
  \bibinfo{pages}{7387} (\bibinfo{year}{1989}).

\bibitem[{\citenamefont{Hasan and Kane}(2010)}]{HasanReview:2010}
\bibinfo{author}{\bibfnamefont{M.~Z.} \bibnamefont{Hasan}} \bibnamefont{and}
  \bibinfo{author}{\bibfnamefont{C.~L.} \bibnamefont{Kane}},
  \bibinfo{journal}{Rev. Mod. Phys.} \textbf{\bibinfo{volume}{82}},
  \bibinfo{pages}{3045} (\bibinfo{year}{2010}).

\bibitem[{\citenamefont{Qi et~al.}(2006)\citenamefont{Qi, Wu, and
  Zhang}}]{Qi:2006}
\bibinfo{author}{\bibfnamefont{X.-L.} \bibnamefont{Qi}},
  \bibinfo{author}{\bibfnamefont{Y.-S.} \bibnamefont{Wu}}, \bibnamefont{and}
  \bibinfo{author}{\bibfnamefont{S.-C.} \bibnamefont{Zhang}},
  \bibinfo{journal}{Phys. Rev. B} \textbf{\bibinfo{volume}{74}},
  \bibinfo{pages}{085308} (\bibinfo{year}{2006}).

\bibitem[{\citenamefont{Chiu et~al.}(2016)\citenamefont{Chiu, Teo, Schnyder,
  and Ryu}}]{ChiuReview:2016}
\bibinfo{author}{\bibfnamefont{C.-K.} \bibnamefont{Chiu}},
  \bibinfo{author}{\bibfnamefont{J.~C.~Y.} \bibnamefont{Teo}},
  \bibinfo{author}{\bibfnamefont{A.~P.} \bibnamefont{Schnyder}},
  \bibnamefont{and} \bibinfo{author}{\bibfnamefont{S.}~\bibnamefont{Ryu}},
  \bibinfo{journal}{Rev. Mod. Phys.} \textbf{\bibinfo{volume}{88}},
  \bibinfo{pages}{035055} (\bibinfo{year}{2016}).

\bibitem[{\citenamefont{Wen}(2017)}]{WenReview:2017}
\bibinfo{author}{\bibfnamefont{X.-G.} \bibnamefont{Wen}},
  \bibinfo{journal}{Rev. Mod. Phys.} \textbf{\bibinfo{volume}{89}},
  \bibinfo{pages}{041004} (\bibinfo{year}{2017}).

\bibitem[{\citenamefont{Halperin}(1982)}]{Halperin:1982}
\bibinfo{author}{\bibfnamefont{B.~I.} \bibnamefont{Halperin}},
  \bibinfo{journal}{Phys. Rev. B} \textbf{\bibinfo{volume}{25}},
  \bibinfo{pages}{2185} (\bibinfo{year}{1982}).

\bibitem[{\citenamefont{Arovas et~al.}(1984)\citenamefont{Arovas, Schrieffer,
  and Wilczek}}]{Arovas:1984}
\bibinfo{author}{\bibfnamefont{D.}~\bibnamefont{Arovas}},
  \bibinfo{author}{\bibfnamefont{J.~R.} \bibnamefont{Schrieffer}},
  \bibnamefont{and} \bibinfo{author}{\bibfnamefont{F.}~\bibnamefont{Wilczek}},
  \bibinfo{journal}{Phys. Rev. Lett.} \textbf{\bibinfo{volume}{53}},
  \bibinfo{pages}{722} (\bibinfo{year}{1984}).

\bibitem[{\citenamefont{Halperin}(1984)}]{Halperin:1984}
\bibinfo{author}{\bibfnamefont{B.~I.} \bibnamefont{Halperin}},
  \bibinfo{journal}{Phys. Rev. Lett.} \textbf{\bibinfo{volume}{52}},
  \bibinfo{pages}{1583} (\bibinfo{year}{1984}).

\bibitem[{\citenamefont{Niu et~al.}(1985)\citenamefont{Niu, Thouless, and
  Wu}}]{Niu85}
\bibinfo{author}{\bibfnamefont{Q.}~\bibnamefont{Niu}},
  \bibinfo{author}{\bibfnamefont{D.~J.} \bibnamefont{Thouless}},
  \bibnamefont{and} \bibinfo{author}{\bibfnamefont{Y.-S.} \bibnamefont{Wu}},
  \bibinfo{journal}{Phys. Rev. B} \textbf{\bibinfo{volume}{31}},
  \bibinfo{pages}{3372} (\bibinfo{year}{1985}),
  \urlprefix\url{https://link.aps.org/doi/10.1103/PhysRevB.31.3372}.

\bibitem[{\citenamefont{Kalmeyer and Laughlin}(1987)}]{Kalmeyer:1987}
\bibinfo{author}{\bibfnamefont{V.}~\bibnamefont{Kalmeyer}} \bibnamefont{and}
  \bibinfo{author}{\bibfnamefont{R.~B.} \bibnamefont{Laughlin}},
  \bibinfo{journal}{Phys. Rev. Lett.} \textbf{\bibinfo{volume}{59}},
  \bibinfo{pages}{2095} (\bibinfo{year}{1987}).

\bibitem[{\citenamefont{Wen et~al.}(1989)\citenamefont{Wen, Wilczek, and
  Zee}}]{Wen:1989-2}
\bibinfo{author}{\bibfnamefont{X.-G.} \bibnamefont{Wen}},
  \bibinfo{author}{\bibfnamefont{F.}~\bibnamefont{Wilczek}}, \bibnamefont{and}
  \bibinfo{author}{\bibfnamefont{A.}~\bibnamefont{Zee}},
  \bibinfo{journal}{Phys. Rev. B} \textbf{\bibinfo{volume}{39}},
  \bibinfo{pages}{11413} (\bibinfo{year}{1989}).

\bibitem[{\citenamefont{Witten}(1989)}]{Witten:1989}
\bibinfo{author}{\bibfnamefont{E.}~\bibnamefont{Witten}},
  \bibinfo{journal}{Commun. Math. Phys.} \textbf{\bibinfo{volume}{121}},
  \bibinfo{pages}{351} (\bibinfo{year}{1989}).

\bibitem[{\citenamefont{Wen}(1991)}]{Wen:1991}
\bibinfo{author}{\bibfnamefont{X.-G.} \bibnamefont{Wen}},
  \bibinfo{journal}{Phys. Rev. B} \textbf{\bibinfo{volume}{43}},
  \bibinfo{pages}{11025} (\bibinfo{year}{1991}).

\bibitem[{\citenamefont{Moore and Read}(1991)}]{Moore:1991}
\bibinfo{author}{\bibfnamefont{G.}~\bibnamefont{Moore}} \bibnamefont{and}
  \bibinfo{author}{\bibfnamefont{N.}~\bibnamefont{Read}},
  \bibinfo{journal}{Nucl. Phys.} \textbf{\bibinfo{volume}{B360}},
  \bibinfo{pages}{362} (\bibinfo{year}{1991}).

\bibitem[{\citenamefont{Wen}(1993)}]{Wen:1993}
\bibinfo{author}{\bibfnamefont{X.-G.} \bibnamefont{Wen}},
  \bibinfo{journal}{Phys. Rev. Lett.} \textbf{\bibinfo{volume}{70}},
  \bibinfo{pages}{355} (\bibinfo{year}{1993}).

\bibitem[{\citenamefont{Wen}(1999)}]{Wen:1999}
\bibinfo{author}{\bibfnamefont{X.-G.} \bibnamefont{Wen}},
  \bibinfo{journal}{Phys. Rev. B} \textbf{\bibinfo{volume}{60}},
  \bibinfo{pages}{8827} (\bibinfo{year}{1999}).

\bibitem[{\citenamefont{Bonderson et~al.}(2011)\citenamefont{Bonderson,
  Gurarie, and Nayak}}]{Bonderson:2011}
\bibinfo{author}{\bibfnamefont{P.}~\bibnamefont{Bonderson}},
  \bibinfo{author}{\bibfnamefont{V.}~\bibnamefont{Gurarie}}, \bibnamefont{and}
  \bibinfo{author}{\bibfnamefont{C.}~\bibnamefont{Nayak}},
  \bibinfo{journal}{Phys. Rev. B} \textbf{\bibinfo{volume}{83}},
  \bibinfo{pages}{075303} (\bibinfo{year}{2011}).

\bibitem[{\citenamefont{Kitaev}(2003)}]{Kitaev:2003}
\bibinfo{author}{\bibfnamefont{A.~Y.} \bibnamefont{Kitaev}},
  \bibinfo{journal}{Ann. Phys. (N.Y.)} \textbf{\bibinfo{volume}{303}},
  \bibinfo{pages}{2} (\bibinfo{year}{2003}).

\bibitem[{\citenamefont{Kitaev and Preskill}(2006)}]{Kitaev:2006}
\bibinfo{author}{\bibfnamefont{A.}~\bibnamefont{Kitaev}} \bibnamefont{and}
  \bibinfo{author}{\bibfnamefont{J.}~\bibnamefont{Preskill}},
  \bibinfo{journal}{Phys. Rev. Lett.} \textbf{\bibinfo{volume}{96}},
  \bibinfo{pages}{110404} (\bibinfo{year}{2006}).

\bibitem[{\citenamefont{Levin and Wen}(2006)}]{Levin:2006}
\bibinfo{author}{\bibfnamefont{M.}~\bibnamefont{Levin}} \bibnamefont{and}
  \bibinfo{author}{\bibfnamefont{X.-G.} \bibnamefont{Wen}},
  \bibinfo{journal}{Phys. Rev. Lett.} \textbf{\bibinfo{volume}{96}},
  \bibinfo{pages}{110405} (\bibinfo{year}{2006}).

\bibitem[{\citenamefont{Chen et~al.}(2010)\citenamefont{Chen, Gu, and
  Wen}}]{XieChen:2010}
\bibinfo{author}{\bibfnamefont{X.}~\bibnamefont{Chen}},
  \bibinfo{author}{\bibfnamefont{Z.-C.} \bibnamefont{Gu}}, \bibnamefont{and}
  \bibinfo{author}{\bibfnamefont{X.-G.} \bibnamefont{Wen}},
  \bibinfo{journal}{Phys. Rev. B} \textbf{\bibinfo{volume}{82}},
  \bibinfo{pages}{155138} (\bibinfo{year}{2010}).

\bibitem[{\citenamefont{Tsui et~al.}(1982)\citenamefont{Tsui, Stormer, and
  Gossard}}]{Tsui:1982}
\bibinfo{author}{\bibfnamefont{D.~C.} \bibnamefont{Tsui}},
  \bibinfo{author}{\bibfnamefont{H.~L.} \bibnamefont{Stormer}},
  \bibnamefont{and} \bibinfo{author}{\bibfnamefont{A.~C.}
  \bibnamefont{Gossard}}, \bibinfo{journal}{Phys. Rev. Lett.} p.
  \bibinfo{pages}{1559} (\bibinfo{year}{1982}).

\bibitem[{\citenamefont{Laughlin}(1983)}]{Laughlin:1983}
\bibinfo{author}{\bibfnamefont{R.~B.} \bibnamefont{Laughlin}},
  \bibinfo{journal}{Phys. Rev. Lett.} p. \bibinfo{pages}{1395}
  (\bibinfo{year}{1983}).

\bibitem[{\citenamefont{Haldane}(1983)}]{Haldane:1983}
\bibinfo{author}{\bibfnamefont{F.~D.~M.} \bibnamefont{Haldane}},
  \bibinfo{journal}{Phys. Rev. Lett.} \textbf{\bibinfo{volume}{50}},
  \bibinfo{pages}{1153} (\bibinfo{year}{1983}).

\bibitem[{\citenamefont{Affleck et~al.}(1988)\citenamefont{Affleck, Kennedy,
  Lieb, and Tasaki}}]{Affleck:1988}
\bibinfo{author}{\bibfnamefont{I.}~\bibnamefont{Affleck}},
  \bibinfo{author}{\bibfnamefont{T.}~\bibnamefont{Kennedy}},
  \bibinfo{author}{\bibfnamefont{E.~H.} \bibnamefont{Lieb}}, \bibnamefont{and}
  \bibinfo{author}{\bibfnamefont{H.}~\bibnamefont{Tasaki}},
  \bibinfo{journal}{Commun. Math. Phys.} \textbf{\bibinfo{volume}{115}},
  \bibinfo{pages}{477} (\bibinfo{year}{1988}).

\bibitem[{\citenamefont{Gu and Wen}(2009)}]{Gu:2009}
\bibinfo{author}{\bibfnamefont{Z.-C.} \bibnamefont{Gu}} \bibnamefont{and}
  \bibinfo{author}{\bibfnamefont{X.-G.} \bibnamefont{Wen}},
  \bibinfo{journal}{Phys. Rev. B} \textbf{\bibinfo{volume}{80}},
  \bibinfo{pages}{155131} (\bibinfo{year}{2009}).

\bibitem[{\citenamefont{Kane and Mele}(2005{\natexlab{a}})}]{Kane-Mele:2005}
\bibinfo{author}{\bibfnamefont{C.~L.} \bibnamefont{Kane}} \bibnamefont{and}
  \bibinfo{author}{\bibfnamefont{E.~J.} \bibnamefont{Mele}},
  \bibinfo{journal}{Phys. Rev. Lett.} \textbf{\bibinfo{volume}{95 (14)}},
  \bibinfo{pages}{146802} (\bibinfo{year}{2005}{\natexlab{a}}).

\bibitem[{\citenamefont{Kane and Mele}(2005{\natexlab{b}})}]{Kane-Mele:2005-2}
\bibinfo{author}{\bibfnamefont{C.~L.} \bibnamefont{Kane}} \bibnamefont{and}
  \bibinfo{author}{\bibfnamefont{E.~J.} \bibnamefont{Mele}},
  \bibinfo{journal}{Phys. Rev. Lett.} \textbf{\bibinfo{volume}{95}},
  \bibinfo{pages}{226801} (\bibinfo{year}{2005}{\natexlab{b}}).

\bibitem[{\citenamefont{Bernevig et~al.}(2006)\citenamefont{Bernevig, Hughes,
  and Zhang}}]{Bernevig:2006}
\bibinfo{author}{\bibfnamefont{B.~A.} \bibnamefont{Bernevig}},
  \bibinfo{author}{\bibfnamefont{T.~L.} \bibnamefont{Hughes}},
  \bibnamefont{and} \bibinfo{author}{\bibfnamefont{S.-C.} \bibnamefont{Zhang}},
  \bibinfo{journal}{Science} \textbf{\bibinfo{volume}{314}},
  \bibinfo{pages}{1757} (\bibinfo{year}{2006}).

\bibitem[{\citenamefont{Xu and Moore}(2006)}]{Xu:2006}
\bibinfo{author}{\bibfnamefont{C.}~\bibnamefont{Xu}} \bibnamefont{and}
  \bibinfo{author}{\bibfnamefont{J.~E.} \bibnamefont{Moore}},
  \bibinfo{journal}{Phys. Rev. B} \textbf{\bibinfo{volume}{73}},
  \bibinfo{pages}{045322} (\bibinfo{year}{2006}).

\bibitem[{\citenamefont{Fu et~al.}(2007)\citenamefont{Fu, Kane, and
  Mele}}]{Fu:2007}
\bibinfo{author}{\bibfnamefont{L.~C.} \bibnamefont{Fu}},
  \bibinfo{author}{\bibfnamefont{L.}~\bibnamefont{Kane}}, \bibnamefont{and}
  \bibinfo{author}{\bibfnamefont{E.~J.} \bibnamefont{Mele}},
  \bibinfo{journal}{Phys. Rev. Lett.} \textbf{\bibinfo{volume}{98}},
  \bibinfo{pages}{106803} (\bibinfo{year}{2007}).

\bibitem[{\citenamefont{Moore and Balents}(2007)}]{Moore:2007}
\bibinfo{author}{\bibfnamefont{J.~E.} \bibnamefont{Moore}} \bibnamefont{and}
  \bibinfo{author}{\bibfnamefont{L.}~\bibnamefont{Balents}},
  \bibinfo{journal}{Phys. Rev. B} \textbf{\bibinfo{volume}{75}},
  \bibinfo{pages}{121306} (\bibinfo{year}{2007}).

\bibitem[{\citenamefont{Qi et~al.}(2008)\citenamefont{Qi, Hughes, and
  Zhang}}]{Qi:2008}
\bibinfo{author}{\bibfnamefont{X.-L.} \bibnamefont{Qi}},
  \bibinfo{author}{\bibfnamefont{T.}~\bibnamefont{Hughes}}, \bibnamefont{and}
  \bibinfo{author}{\bibfnamefont{S.-C.} \bibnamefont{Zhang}},
  \bibinfo{journal}{Phys. Rev. B} \textbf{\bibinfo{volume}{78}},
  \bibinfo{pages}{195424} (\bibinfo{year}{2008}).

\bibitem[{\citenamefont{Schnyder et~al.}(2008)\citenamefont{Schnyder, Ryu,
  Furusaki, and Ludwig}}]{Schnyder:2008}
\bibinfo{author}{\bibfnamefont{A.~P.} \bibnamefont{Schnyder}},
  \bibinfo{author}{\bibfnamefont{S.}~\bibnamefont{Ryu}},
  \bibinfo{author}{\bibfnamefont{A.}~\bibnamefont{Furusaki}}, \bibnamefont{and}
  \bibinfo{author}{\bibfnamefont{A.~W.~W.} \bibnamefont{Ludwig}},
  \bibinfo{journal}{Phys. Rev. B} \textbf{\bibinfo{volume}{78}},
  \bibinfo{pages}{195125} (\bibinfo{year}{2008}).

\bibitem[{\citenamefont{Kitaev}(2009)}]{Kitaev:2009}
\bibinfo{author}{\bibfnamefont{A.}~\bibnamefont{Kitaev}}, in
  \emph{\bibinfo{booktitle}{AIP Conference Proceedings}}
  (\bibinfo{year}{2009}), vol. \bibinfo{volume}{1134(1)}, pp.
  \bibinfo{pages}{22--30},
  \urlprefix\url{https://aip.scitation.org/doi/abs/10.1063/1.3149495}.

\bibitem[{\citenamefont{Bernevig and Hughes}(2013)}]{Bernevig-book}
\bibinfo{author}{\bibfnamefont{B.~A.} \bibnamefont{Bernevig}} \bibnamefont{and}
  \bibinfo{author}{\bibfnamefont{T.~L.} \bibnamefont{Hughes}},
  \emph{\bibinfo{title}{Topological Insulators and Topological
  Superconductors}} (\bibinfo{publisher}{Princeton University Press},
  \bibinfo{year}{2013}), ISBN \bibinfo{isbn}{9780691151755}.

\bibitem[{\citenamefont{Qi and Zhang}(2011)}]{QiReview:2011}
\bibinfo{author}{\bibfnamefont{X.-L.} \bibnamefont{Qi}} \bibnamefont{and}
  \bibinfo{author}{\bibfnamefont{S.-C.} \bibnamefont{Zhang}},
  \bibinfo{journal}{Rev. Mod. Phys.} \textbf{\bibinfo{volume}{83}},
  \bibinfo{pages}{1057} (\bibinfo{year}{2011}).

\bibitem[{\citenamefont{Rachel}(2018)}]{Rachel:2018}
\bibinfo{author}{\bibfnamefont{S.}~\bibnamefont{Rachel}},
  \bibinfo{journal}{Rep. Prog. Phys.} \textbf{\bibinfo{volume}{81}},
  \bibinfo{pages}{116501} (\bibinfo{year}{2018}).

\bibitem[{\citenamefont{Chen}(2018)}]{Chen:2018}
\bibinfo{author}{\bibfnamefont{W.}~\bibnamefont{Chen}}, \bibinfo{journal}{Phys.
  Rev. B} \textbf{\bibinfo{volume}{97}}, \bibinfo{pages}{115130}
  (\bibinfo{year}{2018}).

\bibitem[{\citenamefont{Chen and Sigrist}(2019)}]{Chen-Sigrist-book:2019}
\bibinfo{author}{\bibfnamefont{W.}~\bibnamefont{Chen}} \bibnamefont{and}
  \bibinfo{author}{\bibfnamefont{M.}~\bibnamefont{Sigrist}},
  \emph{\bibinfo{title}{Topological Phase Transitions: Criticality,
  Universality, and Renormalization Group Approach}}
  (\bibinfo{publisher}{Wiley-Scrivener}, \bibinfo{year}{2019}).

\bibitem[{\citenamefont{Molignini et~al.}(2019)\citenamefont{Molignini, Chen,
  and Chitra}}]{MoligniniReview:2019}
\bibinfo{author}{\bibfnamefont{P.}~\bibnamefont{Molignini}},
  \bibinfo{author}{\bibfnamefont{W.}~\bibnamefont{Chen}}, \bibnamefont{and}
  \bibinfo{author}{\bibfnamefont{R.}~\bibnamefont{Chitra}},
  \bibinfo{journal}{Europhys. Lett.} \textbf{\bibinfo{volume}{128}},
  \bibinfo{pages}{36001} (\bibinfo{year}{2019}).

\bibitem[{\citenamefont{Zegarra et~al.}(2019)\citenamefont{Zegarra, Candido,
  Egues, and Chen}}]{Zegarra:2019}
\bibinfo{author}{\bibfnamefont{A.}~\bibnamefont{Zegarra}},
  \bibinfo{author}{\bibfnamefont{D.~R.} \bibnamefont{Candido}},
  \bibinfo{author}{\bibfnamefont{J.~C.} \bibnamefont{Egues}}, \bibnamefont{and}
  \bibinfo{author}{\bibfnamefont{W.}~\bibnamefont{Chen}},
  \bibinfo{journal}{Phys. Rev. B} \textbf{\bibinfo{volume}{100}},
  \bibinfo{pages}{075114} (\bibinfo{year}{2019}),
  \urlprefix\url{https://link.aps.org/doi/10.1103/PhysRevB.100.075114}.

\bibitem[{\citenamefont{Lindner et~al.}(2011)\citenamefont{Lindner, Refael, and
  Galitski}}]{Lindner:2011}
\bibinfo{author}{\bibfnamefont{N.~H.} \bibnamefont{Lindner}},
  \bibinfo{author}{\bibfnamefont{G.}~\bibnamefont{Refael}}, \bibnamefont{and}
  \bibinfo{author}{\bibfnamefont{V.}~\bibnamefont{Galitski}},
  \bibinfo{journal}{Nat. Phys.} \textbf{\bibinfo{volume}{7}},
  \bibinfo{pages}{490} (\bibinfo{year}{2011}).

\bibitem[{\citenamefont{Kitagawa et~al.}(2010)\citenamefont{Kitagawa, Berg,
  Rudner, and Demler}}]{Kitagawa:2010}
\bibinfo{author}{\bibfnamefont{T.}~\bibnamefont{Kitagawa}},
  \bibinfo{author}{\bibfnamefont{E.}~\bibnamefont{Berg}},
  \bibinfo{author}{\bibfnamefont{M.}~\bibnamefont{Rudner}}, \bibnamefont{and}
  \bibinfo{author}{\bibfnamefont{E.}~\bibnamefont{Demler}},
  \bibinfo{journal}{Phys. Rev. B} \textbf{\bibinfo{volume}{82}},
  \bibinfo{pages}{235114} (\bibinfo{year}{2010}).

\bibitem[{\citenamefont{Liu et~al.}(2013)\citenamefont{Liu, Levchenko, and
  Baranger}}]{Liu:2013}
\bibinfo{author}{\bibfnamefont{D.~E.} \bibnamefont{Liu}},
  \bibinfo{author}{\bibfnamefont{A.}~\bibnamefont{Levchenko}},
  \bibnamefont{and} \bibinfo{author}{\bibfnamefont{H.~U.}
  \bibnamefont{Baranger}}, \bibinfo{journal}{Phys. Rev. Lett.}
  \textbf{\bibinfo{volume}{111}}, \bibinfo{pages}{047002}
  (\bibinfo{year}{2013}).

\bibitem[{\citenamefont{Cayssol et~al.}(2013)\citenamefont{Cayssol, D\'{o}ra,
  Simon, and Moessner}}]{Cayssol:2013}
\bibinfo{author}{\bibfnamefont{J.}~\bibnamefont{Cayssol}},
  \bibinfo{author}{\bibfnamefont{B.}~\bibnamefont{D\'{o}ra}},
  \bibinfo{author}{\bibfnamefont{F.}~\bibnamefont{Simon}}, \bibnamefont{and}
  \bibinfo{author}{\bibfnamefont{R.}~\bibnamefont{Moessner}},
  \bibinfo{journal}{Phys. Status Solidi RRL} \textbf{\bibinfo{volume}{7}},
  \bibinfo{pages}{101} (\bibinfo{year}{2013}).

\bibitem[{\citenamefont{Thakurathi et~al.}(2013)\citenamefont{Thakurathi,
  Patel, Sen, and Dutta}}]{Thakurathi:2013}
\bibinfo{author}{\bibfnamefont{M.}~\bibnamefont{Thakurathi}},
  \bibinfo{author}{\bibfnamefont{A.~A.} \bibnamefont{Patel}},
  \bibinfo{author}{\bibfnamefont{D.}~\bibnamefont{Sen}}, \bibnamefont{and}
  \bibinfo{author}{\bibfnamefont{A.}~\bibnamefont{Dutta}},
  \bibinfo{journal}{Phys. Rev. B} \textbf{\bibinfo{volume}{88}},
  \bibinfo{pages}{155133} (\bibinfo{year}{2013}).

\bibitem[{\citenamefont{Graf and Porta}(2013)}]{Graf:2013}
\bibinfo{author}{\bibfnamefont{G.~M.} \bibnamefont{Graf}} \bibnamefont{and}
  \bibinfo{author}{\bibfnamefont{M.}~\bibnamefont{Porta}},
  \bibinfo{journal}{Commun. Math. Phys.} \textbf{\bibinfo{volume}{324 (3)}},
  \bibinfo{pages}{851} (\bibinfo{year}{2013}).

\bibitem[{\citenamefont{Rudner et~al.}(2013)\citenamefont{Rudner, Lindner,
  Berg, and Levin}}]{Rudner:2013}
\bibinfo{author}{\bibfnamefont{M.~S.} \bibnamefont{Rudner}},
  \bibinfo{author}{\bibfnamefont{N.~H.} \bibnamefont{Lindner}},
  \bibinfo{author}{\bibfnamefont{E.}~\bibnamefont{Berg}}, \bibnamefont{and}
  \bibinfo{author}{\bibfnamefont{M.}~\bibnamefont{Levin}},
  \bibinfo{journal}{Phys. Rev. X} \textbf{\bibinfo{volume}{3}},
  \bibinfo{pages}{031005} (\bibinfo{year}{2013}).

\bibitem[{\citenamefont{Farrell and Pereg-Barnea}(2016)}]{Farrell:2016}
\bibinfo{author}{\bibfnamefont{A.}~\bibnamefont{Farrell}} \bibnamefont{and}
  \bibinfo{author}{\bibfnamefont{T.}~\bibnamefont{Pereg-Barnea}},
  \bibinfo{journal}{Phys. Rev. B} \textbf{\bibinfo{volume}{93}},
  \bibinfo{pages}{045121} (\bibinfo{year}{2016}).

\bibitem[{\citenamefont{Harper and Roy}(2017)}]{Harper:2017}
\bibinfo{author}{\bibfnamefont{F.}~\bibnamefont{Harper}} \bibnamefont{and}
  \bibinfo{author}{\bibfnamefont{R.}~\bibnamefont{Roy}},
  \bibinfo{journal}{Phys. Rev. Lett.} \textbf{\bibinfo{volume}{118}},
  \bibinfo{pages}{115301} (\bibinfo{year}{2017}).

\bibitem[{\citenamefont{Roy and Harper}(2017)}]{Roy:2017}
\bibinfo{author}{\bibfnamefont{R.}~\bibnamefont{Roy}} \bibnamefont{and}
  \bibinfo{author}{\bibfnamefont{F.}~\bibnamefont{Harper}},
  \bibinfo{journal}{Phys. Rev. B} \textbf{\bibinfo{volume}{96}},
  \bibinfo{pages}{155118} (\bibinfo{year}{2017}).

\bibitem[{\citenamefont{Yao et~al.}(2017)\citenamefont{Yao, Yan, and
  Wang}}]{Yao:2017}
\bibinfo{author}{\bibfnamefont{S.}~\bibnamefont{Yao}},
  \bibinfo{author}{\bibfnamefont{Z.}~\bibnamefont{Yan}}, \bibnamefont{and}
  \bibinfo{author}{\bibfnamefont{Z.}~\bibnamefont{Wang}},
  \bibinfo{journal}{Phys. Rev. B} \textbf{\bibinfo{volume}{96}},
  \bibinfo{pages}{195303} (\bibinfo{year}{2017}).

\bibitem[{\citenamefont{Molignini et~al.}(2017)\citenamefont{Molignini, van
  Nieuwenburg, and Chitra}}]{Molignini:2017}
\bibinfo{author}{\bibfnamefont{P.}~\bibnamefont{Molignini}},
  \bibinfo{author}{\bibfnamefont{E.}~\bibnamefont{van Nieuwenburg}},
  \bibnamefont{and} \bibinfo{author}{\bibfnamefont{R.}~\bibnamefont{Chitra}},
  \bibinfo{journal}{Phys. Rev. B} \textbf{\bibinfo{volume}{96}},
  \bibinfo{pages}{125144} (\bibinfo{year}{2017}).

\bibitem[{\citenamefont{Molignini et~al.}(2018)\citenamefont{Molignini, Chen,
  and Chitra}}]{Molignini:2018}
\bibinfo{author}{\bibfnamefont{P.}~\bibnamefont{Molignini}},
  \bibinfo{author}{\bibfnamefont{W.}~\bibnamefont{Chen}}, \bibnamefont{and}
  \bibinfo{author}{\bibfnamefont{R.}~\bibnamefont{Chitra}},
  \bibinfo{journal}{Phys. Rev. B} \textbf{\bibinfo{volume}{98}},
  \bibinfo{pages}{125129} (\bibinfo{year}{2018}).

\bibitem[{\citenamefont{Esin et~al.}(2018)\citenamefont{Esin, Rudner, Refael,
  and Lindner}}]{Esin:2018}
\bibinfo{author}{\bibfnamefont{I.}~\bibnamefont{Esin}},
  \bibinfo{author}{\bibfnamefont{M.~S.} \bibnamefont{Rudner}},
  \bibinfo{author}{\bibfnamefont{G.}~\bibnamefont{Refael}}, \bibnamefont{and}
  \bibinfo{author}{\bibfnamefont{N.~H.} \bibnamefont{Lindner}},
  \bibinfo{journal}{Phys. Rev. B} \textbf{\bibinfo{volume}{97}},
  \bibinfo{pages}{245401} (\bibinfo{year}{2018}).

\bibitem[{\citenamefont{Molignini et~al.}(2020)\citenamefont{Molignini, Chen,
  and Chitra}}]{Molignini:2019}
\bibinfo{author}{\bibfnamefont{P.}~\bibnamefont{Molignini}},
  \bibinfo{author}{\bibfnamefont{W.}~\bibnamefont{Chen}}, \bibnamefont{and}
  \bibinfo{author}{\bibfnamefont{R.}~\bibnamefont{Chitra}},
  \bibinfo{journal}{Phys. Rev. B} \textbf{\bibinfo{volume}{101}},
  \bibinfo{pages}{165106} (\bibinfo{year}{2020}),
  \urlprefix\url{https://link.aps.org/doi/10.1103/PhysRevB.101.165106}.

\bibitem[{\citenamefont{Seetharam et~al.}(2019)\citenamefont{Seetharam, Bardyn,
  Lindner, Rudner, and Refael}}]{Seetharam:2019}
\bibinfo{author}{\bibfnamefont{K.~I.} \bibnamefont{Seetharam}},
  \bibinfo{author}{\bibfnamefont{C.-E.} \bibnamefont{Bardyn}},
  \bibinfo{author}{\bibfnamefont{N.~H.} \bibnamefont{Lindner}},
  \bibinfo{author}{\bibfnamefont{M.~S.} \bibnamefont{Rudner}},
  \bibnamefont{and} \bibinfo{author}{\bibfnamefont{G.}~\bibnamefont{Refael}},
  \bibinfo{journal}{Phys. Rev. B} \textbf{\bibinfo{volume}{99}},
  \bibinfo{pages}{014307} (\bibinfo{year}{2019}).

\bibitem[{\citenamefont{Molignini}(2019)}]{Molignini:2020-multifrequency}
\bibinfo{author}{\bibfnamefont{P.}~\bibnamefont{Molignini}},
  \bibinfo{journal}{in preparation}  (\bibinfo{year}{2019}).

\bibitem[{\citenamefont{Rudner and Lindner}(2020)}]{Rudner:2020}
\bibinfo{author}{\bibfnamefont{M.~S.} \bibnamefont{Rudner}} \bibnamefont{and}
  \bibinfo{author}{\bibfnamefont{N.~H.} \bibnamefont{Lindner}},
  \bibinfo{journal}{Nature Reviews Physics} \textbf{\bibinfo{volume}{2}},
  \bibinfo{pages}{229} (\bibinfo{year}{2020}).

\bibitem[{\citenamefont{Harper et~al.}(2020)\citenamefont{Harper, Roy, Rudner,
  and Sondhi}}]{Harper:2020}
\bibinfo{author}{\bibfnamefont{F.}~\bibnamefont{Harper}},
  \bibinfo{author}{\bibfnamefont{R.}~\bibnamefont{Roy}},
  \bibinfo{author}{\bibfnamefont{M.~S.} \bibnamefont{Rudner}},
  \bibnamefont{and} \bibinfo{author}{\bibfnamefont{S.}~\bibnamefont{Sondhi}},
  \bibinfo{journal}{Annual Review of Condensed Matter Physics}
  \textbf{\bibinfo{volume}{11}}, \bibinfo{pages}{345} (\bibinfo{year}{2020}),
  \urlprefix\url{https://doi.org/10.1146/annurev-conmatphys-031218-013721}.

\bibitem[{\citenamefont{Molignini et~al.}(2021)\citenamefont{Molignini,
  Celades, Chen, and Chitra}}]{Molignini:2021}
\bibinfo{author}{\bibfnamefont{P.}~\bibnamefont{Molignini}},
  \bibinfo{author}{\bibfnamefont{A.~G.} \bibnamefont{Celades}},
  \bibinfo{author}{\bibfnamefont{W.}~\bibnamefont{Chen}}, \bibnamefont{and}
  \bibinfo{author}{\bibfnamefont{R.}~\bibnamefont{Chitra}},
  \bibinfo{journal}{Phys. Rev. B} \textbf{\bibinfo{volume}{103}},
  \bibinfo{pages}{184507} (\bibinfo{year}{2021}).

\bibitem[{\citenamefont{McGinley and
  Cooper}(2019{\natexlab{a}})}]{mcginley:PRB2019}
\bibinfo{author}{\bibfnamefont{M.}~\bibnamefont{McGinley}} \bibnamefont{and}
  \bibinfo{author}{\bibfnamefont{N.~R.} \bibnamefont{Cooper}},
  \bibinfo{journal}{Phys. Rev. B} \textbf{\bibinfo{volume}{99}},
  \bibinfo{pages}{075148} (\bibinfo{year}{2019}{\natexlab{a}}).

\bibitem[{\citenamefont{McGinley and
  Cooper}(2019{\natexlab{b}})}]{mcginley:PRR2019}
\bibinfo{author}{\bibfnamefont{M.}~\bibnamefont{McGinley}} \bibnamefont{and}
  \bibinfo{author}{\bibfnamefont{N.~R.} \bibnamefont{Cooper}},
  \bibinfo{journal}{Phys. Rev. Research} \textbf{\bibinfo{volume}{1}},
  \bibinfo{pages}{033204} (\bibinfo{year}{2019}{\natexlab{b}}).

\bibitem[{\citenamefont{Garate}(2013)}]{Garate:2013}
\bibinfo{author}{\bibfnamefont{I.}~\bibnamefont{Garate}},
  \bibinfo{journal}{Phys. Rev. Lett.} \textbf{\bibinfo{volume}{110}},
  \bibinfo{pages}{046402} (\bibinfo{year}{2013}).

\bibitem[{\citenamefont{Bardyn et~al.}(2013)\citenamefont{Bardyn, Baranov,
  Kraus, Rico, Imamoglu, Zoller, and Diehl}}]{Bardyn:2013}
\bibinfo{author}{\bibfnamefont{C.-E.} \bibnamefont{Bardyn}},
  \bibinfo{author}{\bibfnamefont{M.~A.} \bibnamefont{Baranov}},
  \bibinfo{author}{\bibfnamefont{C.~V.} \bibnamefont{Kraus}},
  \bibinfo{author}{\bibfnamefont{E.}~\bibnamefont{Rico}},
  \bibinfo{author}{\bibfnamefont{A.}~\bibnamefont{Imamoglu}},
  \bibinfo{author}{\bibfnamefont{P.}~\bibnamefont{Zoller}}, \bibnamefont{and}
  \bibinfo{author}{\bibfnamefont{S.}~\bibnamefont{Diehl}},
  \bibinfo{journal}{New J. Phys.} \textbf{\bibinfo{volume}{15}},
  \bibinfo{pages}{085001} (\bibinfo{year}{2013}).

\bibitem[{\citenamefont{Saha and Garate}(2014)}]{Saha:2014}
\bibinfo{author}{\bibfnamefont{K.}~\bibnamefont{Saha}} \bibnamefont{and}
  \bibinfo{author}{\bibfnamefont{I.}~\bibnamefont{Garate}},
  \bibinfo{journal}{Phys. Rev. B} \textbf{\bibinfo{volume}{89}},
  \bibinfo{pages}{205103} (\bibinfo{year}{2014}).

\bibitem[{\citenamefont{Saha et~al.}(2015)\citenamefont{Saha, L\'{e}gar\'{e},
  and Garate}}]{Saha:2015}
\bibinfo{author}{\bibfnamefont{K.}~\bibnamefont{Saha}},
  \bibinfo{author}{\bibfnamefont{K.}~\bibnamefont{L\'{e}gar\'{e}}},
  \bibnamefont{and} \bibinfo{author}{\bibfnamefont{I.}~\bibnamefont{Garate}},
  \bibinfo{journal}{Phys. Rev. Lett.} \textbf{\bibinfo{volume}{115}},
  \bibinfo{pages}{176405} (\bibinfo{year}{2015}).

\bibitem[{\citenamefont{Budich and Diehl}(2015)}]{Budich:2015}
\bibinfo{author}{\bibfnamefont{J.~C.} \bibnamefont{Budich}} \bibnamefont{and}
  \bibinfo{author}{\bibfnamefont{S.}~\bibnamefont{Diehl}},
  \bibinfo{journal}{Phys. Rev. B} \textbf{\bibinfo{volume}{91}},
  \bibinfo{pages}{165140} (\bibinfo{year}{2015}).

\bibitem[{\citenamefont{Grusdt}(2017)}]{Grusdt:2017}
\bibinfo{author}{\bibfnamefont{F.}~\bibnamefont{Grusdt}},
  \bibinfo{journal}{Phys. Rev. B} \textbf{\bibinfo{volume}{95}},
  \bibinfo{pages}{075106} (\bibinfo{year}{2017}).

\bibitem[{\citenamefont{Monserrat and Vanderbilt}(2016)}]{Monserrat:2016}
\bibinfo{author}{\bibfnamefont{B.}~\bibnamefont{Monserrat}} \bibnamefont{and}
  \bibinfo{author}{\bibfnamefont{D.}~\bibnamefont{Vanderbilt}},
  \bibinfo{journal}{Phys. Rev. Lett.} \textbf{\bibinfo{volume}{117}},
  \bibinfo{pages}{226801} (\bibinfo{year}{2016}).

\bibitem[{\citenamefont{Bhattacharya et~al.}(2017)\citenamefont{Bhattacharya,
  Bandyopadhyay, and Dutta}}]{Bhattacharya:2017}
\bibinfo{author}{\bibfnamefont{U.}~\bibnamefont{Bhattacharya}},
  \bibinfo{author}{\bibfnamefont{S.}~\bibnamefont{Bandyopadhyay}},
  \bibnamefont{and} \bibinfo{author}{\bibfnamefont{A.}~\bibnamefont{Dutta}},
  \bibinfo{journal}{Phys. Rev. B} \textbf{\bibinfo{volume}{96}},
  \bibinfo{pages}{180303(R)} (\bibinfo{year}{2017}).

\bibitem[{\citenamefont{Bardyn et~al.}(2018)\citenamefont{Bardyn, Wawer,
  Altland, Fleischhauer, and Diehl}}]{Bardyn:2018}
\bibinfo{author}{\bibfnamefont{C.-E.} \bibnamefont{Bardyn}},
  \bibinfo{author}{\bibfnamefont{L.}~\bibnamefont{Wawer}},
  \bibinfo{author}{\bibfnamefont{A.}~\bibnamefont{Altland}},
  \bibinfo{author}{\bibfnamefont{M.}~\bibnamefont{Fleischhauer}},
  \bibnamefont{and} \bibinfo{author}{\bibfnamefont{S.}~\bibnamefont{Diehl}},
  \bibinfo{journal}{Phys. Rev. X} \textbf{\bibinfo{volume}{8}},
  \bibinfo{pages}{011035} (\bibinfo{year}{2018}).

\bibitem[{\citenamefont{Coser and P{\'e}rez-Garc{\'\i}a}(2019)}]{Coser:2019}
\bibinfo{author}{\bibfnamefont{A.}~\bibnamefont{Coser}} \bibnamefont{and}
  \bibinfo{author}{\bibfnamefont{D.}~\bibnamefont{P{\'e}rez-Garc{\'\i}a}},
  \bibinfo{journal}{Quantum} \textbf{\bibinfo{volume}{3}}, \bibinfo{pages}{174}
  (\bibinfo{year}{2019}).

\bibitem[{\citenamefont{induced topological insulators: A no-go theorem and
  a~recipe}(2019)}]{Goldstein:2019}
\bibinfo{author}{\bibfnamefont{D.}~\bibnamefont{induced topological insulators:
  A no-go theorem}} \bibnamefont{and} \bibinfo{author}{\bibnamefont{a~recipe}},
  \bibinfo{journal}{SciPost Phys.} \textbf{\bibinfo{volume}{7}},
  \bibinfo{pages}{067} (\bibinfo{year}{2019}).

\bibitem[{\citenamefont{Lu et~al.}(2020)\citenamefont{Lu, Hsieh, and
  Grover}}]{Lu:2020}
\bibinfo{author}{\bibfnamefont{T.-C.} \bibnamefont{Lu}},
  \bibinfo{author}{\bibfnamefont{T.~H.} \bibnamefont{Hsieh}}, \bibnamefont{and}
  \bibinfo{author}{\bibfnamefont{T.}~\bibnamefont{Grover}},
  \bibinfo{journal}{Phys Rev. Lett.} \textbf{\bibinfo{volume}{125}},
  \bibinfo{pages}{116801} (\bibinfo{year}{2020}).

\bibitem[{\citenamefont{Shapourian et~al.}(2021)\citenamefont{Shapourian, Liu,
  Kudler-Flam, and Vishwanath}}]{Shapourian:2021}
\bibinfo{author}{\bibfnamefont{H.}~\bibnamefont{Shapourian}},
  \bibinfo{author}{\bibfnamefont{S.}~\bibnamefont{Liu}},
  \bibinfo{author}{\bibfnamefont{J.}~\bibnamefont{Kudler-Flam}},
  \bibnamefont{and}
  \bibinfo{author}{\bibfnamefont{A.}~\bibnamefont{Vishwanath}},
  \bibinfo{journal}{Phys. Rev. X Quantum} \textbf{\bibinfo{volume}{2}},
  \bibinfo{pages}{030347} (\bibinfo{year}{2021}).

\bibitem[{\citenamefont{Lieu et~al.}(2020)\citenamefont{Lieu, McGinley, and
  Cooper}}]{Lieu:2020}
\bibinfo{author}{\bibfnamefont{S.}~\bibnamefont{Lieu}},
  \bibinfo{author}{\bibfnamefont{M.}~\bibnamefont{McGinley}}, \bibnamefont{and}
  \bibinfo{author}{\bibfnamefont{N.~R.} \bibnamefont{Cooper}},
  \bibinfo{journal}{Phys. Rev. Lett.} \textbf{\bibinfo{volume}{124}},
  \bibinfo{pages}{040401} (\bibinfo{year}{2020}).

\bibitem[{\citenamefont{Altland et~al.}(2021)\citenamefont{Altland,
  Fleischhauer, and Diehl}}]{Altland:2021}
\bibinfo{author}{\bibfnamefont{A.}~\bibnamefont{Altland}},
  \bibinfo{author}{\bibfnamefont{M.}~\bibnamefont{Fleischhauer}},
  \bibnamefont{and} \bibinfo{author}{\bibfnamefont{S.}~\bibnamefont{Diehl}},
  \bibinfo{journal}{Phys. Rev. X} \textbf{\bibinfo{volume}{11}},
  \bibinfo{pages}{021037} (\bibinfo{year}{2021}).

\bibitem[{\citenamefont{Ashida et~al.}(2020)\citenamefont{Ashida, Gong, and
  Ueda}}]{Ashida-review:2020}
\bibinfo{author}{\bibfnamefont{Y.}~\bibnamefont{Ashida}},
  \bibinfo{author}{\bibfnamefont{Z.}~\bibnamefont{Gong}}, \bibnamefont{and}
  \bibinfo{author}{\bibfnamefont{M.}~\bibnamefont{Ueda}},
  \bibinfo{journal}{Advances in Physics} \textbf{\bibinfo{volume}{69}},
  \bibinfo{pages}{249} (\bibinfo{year}{2020}).

\bibitem[{\citenamefont{Bergholtz et~al.}(2021)\citenamefont{Bergholtz, Budich,
  and Kunst}}]{Bergholtz-Review:2021}
\bibinfo{author}{\bibfnamefont{E.~J.} \bibnamefont{Bergholtz}},
  \bibinfo{author}{\bibfnamefont{J.~C.} \bibnamefont{Budich}},
  \bibnamefont{and} \bibinfo{author}{\bibfnamefont{F.~K.} \bibnamefont{Kunst}},
  \bibinfo{journal}{Rev. Mod. Phys.} \textbf{\bibinfo{volume}{93}},
  \bibinfo{pages}{015005} (\bibinfo{year}{2021}).

\bibitem[{\citenamefont{Gong et~al.}(2018)\citenamefont{Gong, Ashida, Kawabata,
  Takasan, Higashikawa, and Ueda}}]{Gong-nonhermitian:2018}
\bibinfo{author}{\bibfnamefont{Z.}~\bibnamefont{Gong}},
  \bibinfo{author}{\bibfnamefont{Y.}~\bibnamefont{Ashida}},
  \bibinfo{author}{\bibfnamefont{K.}~\bibnamefont{Kawabata}},
  \bibinfo{author}{\bibfnamefont{K.}~\bibnamefont{Takasan}},
  \bibinfo{author}{\bibfnamefont{S.}~\bibnamefont{Higashikawa}},
  \bibnamefont{and} \bibinfo{author}{\bibfnamefont{M.}~\bibnamefont{Ueda}},
  \bibinfo{journal}{Phys. Rev. X} \textbf{\bibinfo{volume}{8}},
  \bibinfo{pages}{031079} (\bibinfo{year}{2018}).

\bibitem[{\citenamefont{Shibata and Katsura}(2019)}]{Shibata:2019}
\bibinfo{author}{\bibfnamefont{N.}~\bibnamefont{Shibata}} \bibnamefont{and}
  \bibinfo{author}{\bibfnamefont{H.}~\bibnamefont{Katsura}},
  \bibinfo{journal}{Phys. Rev. B} \textbf{\bibinfo{volume}{99}},
  \bibinfo{pages}{174303} (\bibinfo{year}{2019}).

\bibitem[{\citenamefont{Minganti et~al.}(2019)\citenamefont{Minganti,
  Miranowicz, Chhajlany, and Nori}}]{Minganti:2019}
\bibinfo{author}{\bibfnamefont{F.}~\bibnamefont{Minganti}},
  \bibinfo{author}{\bibfnamefont{A.}~\bibnamefont{Miranowicz}},
  \bibinfo{author}{\bibfnamefont{R.~W.} \bibnamefont{Chhajlany}},
  \bibnamefont{and} \bibinfo{author}{\bibfnamefont{F.}~\bibnamefont{Nori}},
  \bibinfo{journal}{Phys. Rev. A} \textbf{\bibinfo{volume}{100}},
  \bibinfo{pages}{062131} (\bibinfo{year}{2019}).

\bibitem[{\citenamefont{Rahul and Sarkar}(2022)}]{Rahul:2022}
\bibinfo{author}{\bibfnamefont{S.}~\bibnamefont{Rahul}} \bibnamefont{and}
  \bibinfo{author}{\bibfnamefont{S.}~\bibnamefont{Sarkar}},
  \bibinfo{journal}{Scientific Reports} \textbf{\bibinfo{volume}{12}},
  \bibinfo{pages}{6993} (\bibinfo{year}{2022}).

\bibitem[{\citenamefont{Tsubota et~al.}(2022)\citenamefont{Tsubota, Yang,
  Akagi, and Katsura}}]{Tsubota:2022}
\bibinfo{author}{\bibfnamefont{S.}~\bibnamefont{Tsubota}},
  \bibinfo{author}{\bibfnamefont{H.}~\bibnamefont{Yang}},
  \bibinfo{author}{\bibfnamefont{Y.}~\bibnamefont{Akagi}}, \bibnamefont{and}
  \bibinfo{author}{\bibfnamefont{H.}~\bibnamefont{Katsura}},
  \bibinfo{journal}{Phys. Rev. B} \textbf{\bibinfo{volume}{105}},
  \bibinfo{pages}{L201113} (\bibinfo{year}{2022}).

\bibitem[{\citenamefont{Uhlmann}(1986)}]{Uhlmann:1986}
\bibinfo{author}{\bibfnamefont{A.}~\bibnamefont{Uhlmann}},
  \bibinfo{journal}{Rep. Math. Phys.} \textbf{\bibinfo{volume}{9}},
  \bibinfo{pages}{273} (\bibinfo{year}{1986}).

\bibitem[{\citenamefont{Viyuela
  et~al.}(2014{\natexlab{a}})\citenamefont{Viyuela, Rivas, and
  Martin-Delgado}}]{ViyuelaPRL112:2014}
\bibinfo{author}{\bibfnamefont{O.}~\bibnamefont{Viyuela}},
  \bibinfo{author}{\bibfnamefont{A.}~\bibnamefont{Rivas}}, \bibnamefont{and}
  \bibinfo{author}{\bibfnamefont{M.~A.} \bibnamefont{Martin-Delgado}},
  \bibinfo{journal}{Phys. Rev. Lett.} \textbf{\bibinfo{volume}{112}},
  \bibinfo{pages}{130401} (\bibinfo{year}{2014}{\natexlab{a}}).

\bibitem[{\citenamefont{Viyuela
  et~al.}(2014{\natexlab{b}})\citenamefont{Viyuela, Rivas, and
  Martin-Delgado}}]{ViyuelaPRL113:2014}
\bibinfo{author}{\bibfnamefont{O.}~\bibnamefont{Viyuela}},
  \bibinfo{author}{\bibfnamefont{A.}~\bibnamefont{Rivas}}, \bibnamefont{and}
  \bibinfo{author}{\bibfnamefont{M.~A.} \bibnamefont{Martin-Delgado}},
  \bibinfo{journal}{Phys. Rev. Lett.} \textbf{\bibinfo{volume}{113}},
  \bibinfo{pages}{076408} (\bibinfo{year}{2014}{\natexlab{b}}).

\bibitem[{\citenamefont{Huang and Arovas}(2014)}]{Huang:2014}
\bibinfo{author}{\bibfnamefont{Z.}~\bibnamefont{Huang}} \bibnamefont{and}
  \bibinfo{author}{\bibfnamefont{D.~P.} \bibnamefont{Arovas}},
  \bibinfo{journal}{Phys. Rev. Lett.} \textbf{\bibinfo{volume}{113}},
  \bibinfo{pages}{076407} (\bibinfo{year}{2014}).

\bibitem[{\citenamefont{Kempkes et~al.}(2016)\citenamefont{Kempkes, Quelle, and
  Smith}}]{Kempkes:2016}
\bibinfo{author}{\bibfnamefont{S.~N.} \bibnamefont{Kempkes}},
  \bibinfo{author}{\bibfnamefont{A.}~\bibnamefont{Quelle}}, \bibnamefont{and}
  \bibinfo{author}{\bibfnamefont{C.~M.} \bibnamefont{Smith}},
  \bibinfo{journal}{Scient. Rep.} \textbf{\bibinfo{volume}{6}},
  \bibinfo{pages}{38530} (\bibinfo{year}{2016}).

\bibitem[{\citenamefont{Quelle et~al.}(2016)\citenamefont{Quelle, Cobanera, and
  Smith}}]{Quelle:2016}
\bibinfo{author}{\bibfnamefont{A.}~\bibnamefont{Quelle}},
  \bibinfo{author}{\bibfnamefont{E.}~\bibnamefont{Cobanera}}, \bibnamefont{and}
  \bibinfo{author}{\bibfnamefont{C.~M.} \bibnamefont{Smith}},
  \bibinfo{journal}{Phys. Rev. B} \textbf{\bibinfo{volume}{94}},
  \bibinfo{pages}{075133} (\bibinfo{year}{2016}).

\bibitem[{\citenamefont{Carollo et~al.}(2017)\citenamefont{Carollo, Spagnolo,
  and Valenti}}]{Carollo:2017}
\bibinfo{author}{\bibfnamefont{A.}~\bibnamefont{Carollo}},
  \bibinfo{author}{\bibfnamefont{B.}~\bibnamefont{Spagnolo}}, \bibnamefont{and}
  \bibinfo{author}{\bibfnamefont{D.}~\bibnamefont{Valenti}},
  \bibinfo{journal}{Scientific Reports} \textbf{\bibinfo{volume}{8}},
  \bibinfo{pages}{10} (\bibinfo{year}{2017}).

\bibitem[{\citenamefont{Viyuela et~al.}(2018)\citenamefont{Viyuela, Rivas,
  Gasparinetti, Wallraff, Filipp, and Martin-Delgado}}]{Viyuela:2018}
\bibinfo{author}{\bibfnamefont{O.}~\bibnamefont{Viyuela}},
  \bibinfo{author}{\bibfnamefont{A.}~\bibnamefont{Rivas}},
  \bibinfo{author}{\bibfnamefont{S.}~\bibnamefont{Gasparinetti}},
  \bibinfo{author}{\bibfnamefont{A.}~\bibnamefont{Wallraff}},
  \bibinfo{author}{\bibfnamefont{S.}~\bibnamefont{Filipp}}, \bibnamefont{and}
  \bibinfo{author}{\bibfnamefont{M.~A.} \bibnamefont{Martin-Delgado}},
  \bibinfo{journal}{npj Quantum Information} \textbf{\bibinfo{volume}{4}},
  \bibinfo{pages}{10} (\bibinfo{year}{2018}).

\bibitem[{\citenamefont{Simon}(1983)}]{Barry:1983}
\bibinfo{author}{\bibfnamefont{B.}~\bibnamefont{Simon}},
  \bibinfo{journal}{Phys. Rev. Lett.} \textbf{\bibinfo{volume}{51}},
  \bibinfo{pages}{2167} (\bibinfo{year}{1983}).

\bibitem[{\citenamefont{Berry}(1984)}]{Berry:1984}
\bibinfo{author}{\bibfnamefont{M.~V.} \bibnamefont{Berry}},
  \bibinfo{journal}{Proc. R. Soc. A} \textbf{\bibinfo{volume}{392}},
  \bibinfo{pages}{45} (\bibinfo{year}{1984}).

\bibitem[{\citenamefont{Wilczek and Zee}(1984)}]{Wilczek:1984}
\bibinfo{author}{\bibfnamefont{F.}~\bibnamefont{Wilczek}} \bibnamefont{and}
  \bibinfo{author}{\bibfnamefont{A.}~\bibnamefont{Zee}},
  \bibinfo{journal}{Phys. Rev. Lett.} \textbf{\bibinfo{volume}{52}},
  \bibinfo{pages}{2111} (\bibinfo{year}{1984}).

\bibitem[{\citenamefont{Wawer and
  Fleischhauer}(2021{\natexlab{a}})}]{Wawer:2021-2}
\bibinfo{author}{\bibfnamefont{L.}~\bibnamefont{Wawer}} \bibnamefont{and}
  \bibinfo{author}{\bibfnamefont{M.}~\bibnamefont{Fleischhauer}},
  \bibinfo{journal}{Phys. Rev. B} \textbf{\bibinfo{volume}{104}},
  \bibinfo{pages}{094104} (\bibinfo{year}{2021}{\natexlab{a}}).

\bibitem[{\citenamefont{Linzner et~al.}(2016)\citenamefont{Linzner, Wawer,
  Grusdt, and Fleischhauer}}]{Linzner:2016}
\bibinfo{author}{\bibfnamefont{D.}~\bibnamefont{Linzner}},
  \bibinfo{author}{\bibfnamefont{L.}~\bibnamefont{Wawer}},
  \bibinfo{author}{\bibfnamefont{F.}~\bibnamefont{Grusdt}}, \bibnamefont{and}
  \bibinfo{author}{\bibfnamefont{M.}~\bibnamefont{Fleischhauer}},
  \bibinfo{journal}{Phys. Rev. B} \textbf{\bibinfo{volume}{94}},
  \bibinfo{pages}{201105(R)} (\bibinfo{year}{2016}).

\bibitem[{\citenamefont{Mink et~al.}(2019)\citenamefont{Mink, Fleischhauer, and
  Unanyan}}]{Mink:2019}
\bibinfo{author}{\bibfnamefont{C.~D.} \bibnamefont{Mink}},
  \bibinfo{author}{\bibfnamefont{M.}~\bibnamefont{Fleischhauer}},
  \bibnamefont{and} \bibinfo{author}{\bibfnamefont{R.}~\bibnamefont{Unanyan}},
  \bibinfo{journal}{Phys. Rev. B} \textbf{\bibinfo{volume}{100}},
  \bibinfo{pages}{014305} (\bibinfo{year}{2019}).

\bibitem[{\citenamefont{Unanyan et~al.}(2020)\citenamefont{Unanyan,
  Kiefer-Emmanouilidis, and Fleischhauer}}]{Unanyan:2020}
\bibinfo{author}{\bibfnamefont{R.}~\bibnamefont{Unanyan}},
  \bibinfo{author}{\bibfnamefont{M.}~\bibnamefont{Kiefer-Emmanouilidis}},
  \bibnamefont{and}
  \bibinfo{author}{\bibfnamefont{M.}~\bibnamefont{Fleischhauer}},
  \bibinfo{journal}{Phys. Rev. Lett.} \textbf{\bibinfo{volume}{125}},
  \bibinfo{pages}{215701} (\bibinfo{year}{2020}).

\bibitem[{\citenamefont{Wawer et~al.}(2021)\citenamefont{Wawer, Li, and
  Fleischhauer}}]{Wawer:2021-1}
\bibinfo{author}{\bibfnamefont{L.}~\bibnamefont{Wawer}},
  \bibinfo{author}{\bibfnamefont{R.}~\bibnamefont{Li}}, \bibnamefont{and}
  \bibinfo{author}{\bibfnamefont{M.}~\bibnamefont{Fleischhauer}},
  \bibinfo{journal}{Phys. Rev. A} \textbf{\bibinfo{volume}{104}},
  \bibinfo{pages}{012209} (\bibinfo{year}{2021}).

\bibitem[{\citenamefont{Wawer and
  Fleischhauer}(2021{\natexlab{b}})}]{Wawer:2021-3}
\bibinfo{author}{\bibfnamefont{L.}~\bibnamefont{Wawer}} \bibnamefont{and}
  \bibinfo{author}{\bibfnamefont{M.}~\bibnamefont{Fleischhauer}},
  \bibinfo{journal}{Phys. Rev. B} \textbf{\bibinfo{volume}{104}},
  \bibinfo{pages}{214107} (\bibinfo{year}{2021}{\natexlab{b}}).

\bibitem[{\citenamefont{Wawer et~al.}(2022)\citenamefont{Wawer, Unanyan, and
  Fleischhauer}}]{Wawer:2022}
\bibinfo{author}{\bibfnamefont{L.}~\bibnamefont{Wawer}},
  \bibinfo{author}{\bibfnamefont{R.}~\bibnamefont{Unanyan}}, \bibnamefont{and}
  \bibinfo{author}{\bibfnamefont{M.}~\bibnamefont{Fleischhauer}},
  \bibinfo{journal}{arxiv:2110.12280}  (\bibinfo{year}{2022}).

\bibitem[{\citenamefont{Huang et~al.}(2021)\citenamefont{Huang, Sun, and
  Diehl}}]{Huang:2022}
\bibinfo{author}{\bibfnamefont{Z.-M.} \bibnamefont{Huang}},
  \bibinfo{author}{\bibfnamefont{X.-Q.} \bibnamefont{Sun}}, \bibnamefont{and}
  \bibinfo{author}{\bibfnamefont{S.}~\bibnamefont{Diehl}},
  \bibinfo{journal}{arXiv:2109.06891}  (\bibinfo{year}{2021}).

\bibitem[{\citenamefont{Resta}(1998)}]{Resta:1998}
\bibinfo{author}{\bibfnamefont{R.}~\bibnamefont{Resta}}, \bibinfo{journal}{PRL}
  \textbf{\bibinfo{volume}{80}}, \bibinfo{pages}{1800} (\bibinfo{year}{1998}).

\bibitem[{\citenamefont{Su et~al.}(1979)\citenamefont{Su, Schrieffer, and
  Heeger}}]{SSH:1979}
\bibinfo{author}{\bibfnamefont{W.~P.} \bibnamefont{Su}},
  \bibinfo{author}{\bibfnamefont{J.~R.} \bibnamefont{Schrieffer}},
  \bibnamefont{and} \bibinfo{author}{\bibfnamefont{A.~J.}
  \bibnamefont{Heeger}}, \bibinfo{journal}{Phys. Rev. Lett.}
  \textbf{\bibinfo{volume}{42}}, \bibinfo{pages}{1698} (\bibinfo{year}{1979}).

\bibitem[{\citenamefont{Heeger et~al.}(1988)\citenamefont{Heeger, Kivelson,
  Schrieffer, and Su}}]{Heeger:1988}
\bibinfo{author}{\bibfnamefont{A.}~\bibnamefont{Heeger}},
  \bibinfo{author}{\bibfnamefont{S.}~\bibnamefont{Kivelson}},
  \bibinfo{author}{\bibfnamefont{J.}~\bibnamefont{Schrieffer}},
  \bibnamefont{and} \bibinfo{author}{\bibfnamefont{W.}~\bibnamefont{Su}},
  \bibinfo{journal}{Rev. Mod. Phys.} \textbf{\bibinfo{volume}{60}},
  \bibinfo{pages}{781} (\bibinfo{year}{1988}).

\bibitem[{\citenamefont{Asb{\'o}th et~al.}(2016)\citenamefont{Asb{\'o}th,
  Oroszl{\'a}ny, and P{\'a}lyi}}]{Asboth-book}
\bibinfo{author}{\bibfnamefont{J.~K.} \bibnamefont{Asb{\'o}th}},
  \bibinfo{author}{\bibfnamefont{L.}~\bibnamefont{Oroszl{\'a}ny}},
  \bibnamefont{and}
  \bibinfo{author}{\bibfnamefont{A.}~\bibnamefont{P{\'a}lyi}},
  \emph{\bibinfo{title}{A Short Course on Topological Insulators - Band
  Structure and Edge States in One and Two Dimensions}}, vol.
  \bibinfo{volume}{919} of \emph{\bibinfo{series}{Lecture Notes in Physics}}
  (\bibinfo{publisher}{Springer Verlag}, \bibinfo{year}{2016}).

\bibitem[{\citenamefont{B\"{o}hling et~al.}(2018)\citenamefont{B\"{o}hling,
  Engelhardt, Platero, and Schaller}}]{Boehling:2018}
\bibinfo{author}{\bibfnamefont{S.}~\bibnamefont{B\"{o}hling}},
  \bibinfo{author}{\bibfnamefont{G.}~\bibnamefont{Engelhardt}},
  \bibinfo{author}{\bibfnamefont{G.}~\bibnamefont{Platero}}, \bibnamefont{and}
  \bibinfo{author}{\bibfnamefont{G.}~\bibnamefont{Schaller}},
  \bibinfo{journal}{Phys. Rev. B} \textbf{\bibinfo{volume}{98}},
  \bibinfo{pages}{035132} (\bibinfo{year}{2018}).

\bibitem[{\citenamefont{Boross et~al.}(2019)\citenamefont{Boross, Asb{\'o}th,
  Sz{\'e}chenyi, Oroszl{\'a}ny, and P{\'a}lyi}}]{Boross:2019}
\bibinfo{author}{\bibfnamefont{P.}~\bibnamefont{Boross}},
  \bibinfo{author}{\bibfnamefont{J.~K.} \bibnamefont{Asb{\'o}th}},
  \bibinfo{author}{\bibfnamefont{G.}~\bibnamefont{Sz{\'e}chenyi}},
  \bibinfo{author}{\bibfnamefont{L.}~\bibnamefont{Oroszl{\'a}ny}},
  \bibnamefont{and}
  \bibinfo{author}{\bibfnamefont{A.}~\bibnamefont{P{\'a}lyi}},
  \bibinfo{journal}{Phys. Rev. B} \textbf{\bibinfo{volume}{100}},
  \bibinfo{pages}{045414} (\bibinfo{year}{2019}).

\bibitem[{\citenamefont{D\'Angelis et~al.}(2020)\citenamefont{D\'Angelis,
  Pinheiro, Gu'{e}ry-Odelin, Longhi, and Impens}}]{Dangelis:2020}
\bibinfo{author}{\bibfnamefont{F.~M.} \bibnamefont{D\'Angelis}},
  \bibinfo{author}{\bibfnamefont{F.~A.} \bibnamefont{Pinheiro}},
  \bibinfo{author}{\bibfnamefont{D.}~\bibnamefont{Gu'{e}ry-Odelin}},
  \bibinfo{author}{\bibfnamefont{S.}~\bibnamefont{Longhi}}, \bibnamefont{and}
  \bibinfo{author}{\bibfnamefont{F.}~\bibnamefont{Impens}},
  \bibinfo{journal}{Phys. Rev. Research} \textbf{\bibinfo{volume}{2}},
  \bibinfo{pages}{033475} (\bibinfo{year}{2020}).

\bibitem[{\citenamefont{Go et~al.}(2020)\citenamefont{Go, Hong, Lee, Kim, and
  Lee}}]{Go:2020}
\bibinfo{author}{\bibfnamefont{G.}~\bibnamefont{Go}},
  \bibinfo{author}{\bibfnamefont{I.-S.} \bibnamefont{Hong}},
  \bibinfo{author}{\bibfnamefont{S.-W.} \bibnamefont{Lee}},
  \bibinfo{author}{\bibfnamefont{S.~K.} \bibnamefont{Kim}}, \bibnamefont{and}
  \bibinfo{author}{\bibfnamefont{K.-J.} \bibnamefont{Lee}},
  \bibinfo{journal}{Phys. Rev. B} \textbf{\bibinfo{volume}{101}},
  \bibinfo{pages}{134423} (\bibinfo{year}{2020}).

\bibitem[{\citenamefont{de~L\'{e}s\'{e}luc
  et~al.}(2019)\citenamefont{de~L\'{e}s\'{e}luc, Lienhard, Scholl, Barredo,
  Weber, Lang, B\"{u}chler, Lahaye, and Browaeys}}]{Leseluc:2019}
\bibinfo{author}{\bibfnamefont{S.}~\bibnamefont{de~L\'{e}s\'{e}luc}},
  \bibinfo{author}{\bibfnamefont{V.}~\bibnamefont{Lienhard}},
  \bibinfo{author}{\bibfnamefont{P.}~\bibnamefont{Scholl}},
  \bibinfo{author}{\bibfnamefont{D.}~\bibnamefont{Barredo}},
  \bibinfo{author}{\bibfnamefont{S.}~\bibnamefont{Weber}},
  \bibinfo{author}{\bibfnamefont{N.}~\bibnamefont{Lang}},
  \bibinfo{author}{\bibfnamefont{H.~P.} \bibnamefont{B\"{u}chler}},
  \bibinfo{author}{\bibfnamefont{T.}~\bibnamefont{Lahaye}}, \bibnamefont{and}
  \bibinfo{author}{\bibfnamefont{A.}~\bibnamefont{Browaeys}},
  \bibinfo{journal}{Science} \textbf{\bibinfo{volume}{365}},
  \bibinfo{pages}{775} (\bibinfo{year}{2019}).

\bibitem[{\citenamefont{Kanungo et~al.}(2022)\citenamefont{Kanungo, Whalen, Lu,
  Yuan, Dasgupta, Dunning, Hazzard, and Killian}}]{Kanungo:2022}
\bibinfo{author}{\bibfnamefont{S.~K.} \bibnamefont{Kanungo}},
  \bibinfo{author}{\bibfnamefont{J.~D.} \bibnamefont{Whalen}},
  \bibinfo{author}{\bibfnamefont{Y.}~\bibnamefont{Lu}},
  \bibinfo{author}{\bibfnamefont{M.}~\bibnamefont{Yuan}},
  \bibinfo{author}{\bibfnamefont{S.}~\bibnamefont{Dasgupta}},
  \bibinfo{author}{\bibfnamefont{F.~B.} \bibnamefont{Dunning}},
  \bibinfo{author}{\bibfnamefont{K.~R.~A.} \bibnamefont{Hazzard}},
  \bibnamefont{and} \bibinfo{author}{\bibfnamefont{T.~C.}
  \bibnamefont{Killian}}, \bibinfo{journal}{Nature Communications}
  \textbf{\bibinfo{volume}{13}}, \bibinfo{pages}{972} (\bibinfo{year}{2022}).

\bibitem[{\citenamefont{Atala et~al.}(2013)\citenamefont{Atala, Aidelsburger,
  Barreiro, Abanin, Kitagawa, Demler, and Bloch}}]{Atala:2013}
\bibinfo{author}{\bibfnamefont{M.}~\bibnamefont{Atala}},
  \bibinfo{author}{\bibfnamefont{M.}~\bibnamefont{Aidelsburger}},
  \bibinfo{author}{\bibfnamefont{J.~T.} \bibnamefont{Barreiro}},
  \bibinfo{author}{\bibfnamefont{D.}~\bibnamefont{Abanin}},
  \bibinfo{author}{\bibfnamefont{T.}~\bibnamefont{Kitagawa}},
  \bibinfo{author}{\bibfnamefont{E.}~\bibnamefont{Demler}}, \bibnamefont{and}
  \bibinfo{author}{\bibfnamefont{I.}~\bibnamefont{Bloch}},
  \bibinfo{journal}{Nature Physics} \textbf{\bibinfo{volume}{9}},
  \bibinfo{pages}{795} (\bibinfo{year}{2013}).

\bibitem[{\citenamefont{Meier et~al.}(2016)\citenamefont{Meier, Fangzhao, and
  Gadway}}]{Meier:2016}
\bibinfo{author}{\bibfnamefont{E.~J.} \bibnamefont{Meier}},
  \bibinfo{author}{\bibfnamefont{A.~A.} \bibnamefont{Fangzhao}},
  \bibnamefont{and} \bibinfo{author}{\bibfnamefont{B.}~\bibnamefont{Gadway}},
  \bibinfo{journal}{Nature Communications} \textbf{\bibinfo{volume}{7}},
  \bibinfo{pages}{13986} (\bibinfo{year}{2016}).

\bibitem[{\citenamefont{Zhang et~al.}(2018)\citenamefont{Zhang, Zhu, Zhao, Yan,
  and Zhu}}]{Zhang:2018:AiP}
\bibinfo{author}{\bibfnamefont{D.-W.} \bibnamefont{Zhang}},
  \bibinfo{author}{\bibfnamefont{Y.-Q.} \bibnamefont{Zhu}},
  \bibinfo{author}{\bibfnamefont{Y.~X.} \bibnamefont{Zhao}},
  \bibinfo{author}{\bibfnamefont{H.}~\bibnamefont{Yan}}, \bibnamefont{and}
  \bibinfo{author}{\bibfnamefont{S.-L.} \bibnamefont{Zhu}},
  \bibinfo{journal}{Advances in Physics} \textbf{\bibinfo{volume}{67}},
  \bibinfo{pages}{253} (\bibinfo{year}{2018}).

\bibitem[{\citenamefont{Cooper et~al.}(2019)\citenamefont{Cooper, Dalibard, and
  Spielman}}]{CooperReview:2019}
\bibinfo{author}{\bibfnamefont{N.~R.} \bibnamefont{Cooper}},
  \bibinfo{author}{\bibfnamefont{J.}~\bibnamefont{Dalibard}}, \bibnamefont{and}
  \bibinfo{author}{\bibfnamefont{I.~B.} \bibnamefont{Spielman}},
  \bibinfo{journal}{Rev. Mod. Phys.} \textbf{\bibinfo{volume}{91}},
  \bibinfo{pages}{015005} (\bibinfo{year}{2019}).

\bibitem[{\citenamefont{St-Jean et~al.}(2017)\citenamefont{St-Jean, Goblot,
  Galopin, Lema{\^\i}tre, Ozawa, Gratiet, Sagnes, Bloch, and
  Amo}}]{St-Jean:2017}
\bibinfo{author}{\bibfnamefont{P.}~\bibnamefont{St-Jean}},
  \bibinfo{author}{\bibfnamefont{V.}~\bibnamefont{Goblot}},
  \bibinfo{author}{\bibfnamefont{E.}~\bibnamefont{Galopin}},
  \bibinfo{author}{\bibfnamefont{A.}~\bibnamefont{Lema{\^\i}tre}},
  \bibinfo{author}{\bibfnamefont{T.}~\bibnamefont{Ozawa}},
  \bibinfo{author}{\bibfnamefont{L.~L.} \bibnamefont{Gratiet}},
  \bibinfo{author}{\bibfnamefont{I.}~\bibnamefont{Sagnes}},
  \bibinfo{author}{\bibfnamefont{J.}~\bibnamefont{Bloch}}, \bibnamefont{and}
  \bibinfo{author}{\bibfnamefont{A.}~\bibnamefont{Amo}},
  \bibinfo{journal}{Nature Photon} \textbf{\bibinfo{volume}{11}},
  \bibinfo{pages}{651} (\bibinfo{year}{2017}).

\bibitem[{\citenamefont{Youssefi et~al.}(2021)\citenamefont{Youssefi, Bancora,
  Kono, Chegnizadeh, Vovk, Pan, and Kippenberg}}]{Youssefi:2021}
\bibinfo{author}{\bibfnamefont{A.}~\bibnamefont{Youssefi}},
  \bibinfo{author}{\bibfnamefont{A.}~\bibnamefont{Bancora}},
  \bibinfo{author}{\bibfnamefont{S.}~\bibnamefont{Kono}},
  \bibinfo{author}{\bibfnamefont{M.}~\bibnamefont{Chegnizadeh}},
  \bibinfo{author}{\bibfnamefont{T.}~\bibnamefont{Vovk}},
  \bibinfo{author}{\bibfnamefont{J.}~\bibnamefont{Pan}}, \bibnamefont{and}
  \bibinfo{author}{\bibfnamefont{T.~J.} \bibnamefont{Kippenberg}},
  \bibinfo{journal}{arxiv:2111.09133}  (\bibinfo{year}{2021}).

\bibitem[{\citenamefont{J.}(1989)}]{Zak:1989}
\bibinfo{author}{\bibfnamefont{Z.}~\bibnamefont{J.}}, \bibinfo{journal}{Phys.
  Rev. Lett.} \textbf{\bibinfo{volume}{62}}, \bibinfo{pages}{2747}
  (\bibinfo{year}{1989}).

\bibitem[{\citenamefont{Rhim et~al.}(2017)\citenamefont{Rhim, Behrends, and
  Bardarson}}]{Rhim:2017}
\bibinfo{author}{\bibfnamefont{J.~W.} \bibnamefont{Rhim}},
  \bibinfo{author}{\bibfnamefont{J.}~\bibnamefont{Behrends}}, \bibnamefont{and}
  \bibinfo{author}{\bibfnamefont{J.~H.} \bibnamefont{Bardarson}},
  \bibinfo{journal}{Phys. Rev. B} \textbf{\bibinfo{volume}{95}},
  \bibinfo{pages}{035421} (\bibinfo{year}{2017}).

\bibitem[{\citenamefont{Redfield}(1957)}]{Redfield:1957}
\bibinfo{author}{\bibfnamefont{A.~G.} \bibnamefont{Redfield}},
  \bibinfo{journal}{IBM Journal of Research and Development}
  \textbf{\bibinfo{volume}{1}}, \bibinfo{pages}{19} (\bibinfo{year}{1957}).

\bibitem[{\citenamefont{Petruccione and
  Breuer}(2002)}]{Breuer-Petruccione-book}
\bibinfo{author}{\bibfnamefont{F.}~\bibnamefont{Petruccione}} \bibnamefont{and}
  \bibinfo{author}{\bibfnamefont{H.-P.} \bibnamefont{Breuer}},
  \emph{\bibinfo{title}{The Theory of Open Quantum Systems}}
  (\bibinfo{publisher}{Oxford University Press}, \bibinfo{year}{2002}).

\bibitem[{\citenamefont{Prosen and Zunkovic}(2010)}]{Prosen:2010}
\bibinfo{author}{\bibfnamefont{T.}~\bibnamefont{Prosen}} \bibnamefont{and}
  \bibinfo{author}{\bibfnamefont{B.}~\bibnamefont{Zunkovic}},
  \bibinfo{journal}{New Journal of Physics} \textbf{\bibinfo{volume}{12}},
  \bibinfo{pages}{025016} (\bibinfo{year}{2010}).

\bibitem[{\citenamefont{Mozgunov and Lidar}(2020)}]{Mozgunov:2020}
\bibinfo{author}{\bibfnamefont{E.}~\bibnamefont{Mozgunov}} \bibnamefont{and}
  \bibinfo{author}{\bibfnamefont{D.}~\bibnamefont{Lidar}},
  \bibinfo{journal}{Quantum} \textbf{\bibinfo{volume}{4}}, \bibinfo{pages}{227}
  (\bibinfo{year}{2020}).

\bibitem[{\citenamefont{Prosen and Pizorn}(2008)}]{Prosen:2008}
\bibinfo{author}{\bibfnamefont{T.}~\bibnamefont{Prosen}} \bibnamefont{and}
  \bibinfo{author}{\bibfnamefont{I.}~\bibnamefont{Pizorn}},
  \bibinfo{journal}{Phys. Rev. Lett.} \textbf{\bibinfo{volume}{101}},
  \bibinfo{pages}{105701} (\bibinfo{year}{2008}).

\bibitem[{\citenamefont{Rice and Mele}(1982)}]{Rice:1982}
\bibinfo{author}{\bibfnamefont{M.~J.} \bibnamefont{Rice}} \bibnamefont{and}
  \bibinfo{author}{\bibfnamefont{E.~J.} \bibnamefont{Mele}},
  \bibinfo{journal}{Phys. Rev. Lett.} \textbf{\bibinfo{volume}{49}},
  \bibinfo{pages}{1455} (\bibinfo{year}{1982}).

\bibitem[{\citenamefont{Thouless}(1983)}]{Thouless:1983}
\bibinfo{author}{\bibfnamefont{D.~J.} \bibnamefont{Thouless}},
  \bibinfo{journal}{Phys. Rev. B} \textbf{\bibinfo{volume}{27}},
  \bibinfo{pages}{6083} (\bibinfo{year}{1983}).

\bibitem[{\citenamefont{Bravyi}(2005)}]{Bravyi:2005}
\bibinfo{author}{\bibfnamefont{S.}~\bibnamefont{Bravyi}},
  \bibinfo{journal}{Quantum Inf. and Comp.} \textbf{\bibinfo{volume}{5}},
  \bibinfo{pages}{216} (\bibinfo{year}{2005}).

\bibitem[{\citenamefont{Lieu et~al.}(2022)\citenamefont{Lieu, McGinley,
  Shtanko, Cooper, and Gorshkov}}]{Lieu:2022}
\bibinfo{author}{\bibfnamefont{S.}~\bibnamefont{Lieu}},
  \bibinfo{author}{\bibfnamefont{M.}~\bibnamefont{McGinley}},
  \bibinfo{author}{\bibfnamefont{O.}~\bibnamefont{Shtanko}},
  \bibinfo{author}{\bibfnamefont{N.~R.} \bibnamefont{Cooper}},
  \bibnamefont{and} \bibinfo{author}{\bibfnamefont{A.~V.}
  \bibnamefont{Gorshkov}}, \bibinfo{journal}{Phys. Rev. B}
  \textbf{\bibinfo{volume}{105}}, \bibinfo{pages}{L121104}
  (\bibinfo{year}{2022}).

\bibitem[{\citenamefont{Dangel}(2017)}]{Dangel-thesis:2017}
\bibinfo{author}{\bibfnamefont{F.}~\bibnamefont{Dangel}}, Master's thesis,
  \bibinfo{school}{University of Stuttgart, Germany} (\bibinfo{year}{2017}).

\bibitem[{\citenamefont{Dangel et~al.}(2018)\citenamefont{Dangel, Wagner,
  Cartarius, Main, and Wunner}}]{Dangel:2018}
\bibinfo{author}{\bibfnamefont{F.}~\bibnamefont{Dangel}},
  \bibinfo{author}{\bibfnamefont{M.}~\bibnamefont{Wagner}},
  \bibinfo{author}{\bibfnamefont{H.}~\bibnamefont{Cartarius}},
  \bibinfo{author}{\bibfnamefont{J.}~\bibnamefont{Main}}, \bibnamefont{and}
  \bibinfo{author}{\bibfnamefont{G.}~\bibnamefont{Wunner}},
  \bibinfo{journal}{Phys. Rev. A} \textbf{\bibinfo{volume}{98}},
  \bibinfo{pages}{013628} (\bibinfo{year}{2018}).

\end{thebibliography}

\end{document}